\documentclass[11pt]{article}

\usepackage[utf8]{inputenc}   
\usepackage{amsmath, amssymb} 
\usepackage{graphicx}         
\usepackage{geometry}         
\usepackage{booktabs}         
\usepackage{tabularx}         
\usepackage{adjustbox}        
\usepackage{multirow}         
\usepackage{caption, subcaption} 
\usepackage[linesnumbered,ruled,vlined]{algorithm2e} 
\usepackage{titlesec}         
\usepackage{enumitem}         
\usepackage{pdflscape}        
\usepackage{hyperref}         
\usepackage{longtable}        

\newcounter{paracount}[subsubsection]
\renewcommand{\theparacount}{\alph{paracount}}
\makeatletter
\renewcommand{\paragraph}{\@startsection{paragraph}{4}{\z@}%
	{-3.25ex \@plus -1ex \@minus -0.2ex}%
	{1.5ex \@plus 0.2ex}%
	{\normalfont\normalsize\stepcounter{paracount}\theparacount)\hspace{0.5em}}}
\makeatother
\titleformat{\paragraph}[runin]{\normalfont\normalsize}{}{0em}{}[:]
\graphicspath{ {./Figures/} }

\title{InTec: integrated things‑edge computing\\ A framework
	for distributing machine learning pipelines in edge AI
	systems}
\author{Habib Larian$^{1,2}$ \\
	Faramarz Safi-Esfahani$^{1,2,3}$ \\
	\vspace{0.5em}
	\small $^1$ Faculty of Computer Engineering, Najafabad Branch, Islamic Azad University, Najafabad, Iran \\
	\small $^2$ Big Data Research Center, Najafabad Branch, Islamic Azad University, Najafabad, Iran\\
	\small $^3$ School of Information Systems and Modelling, University of Technology Sydney, Ultimo, Sydney, NSW, Australia}
\date{Published: 31 December 2024}

\begin{document}
	
	\maketitle	
	\begin{abstract}
		With the rapid expansion of the Internet of Things (IoT), sensors, smartphones, and
		wearables have become integral to daily life, powering smart applications in home
		automation, healthcare, and intelligent transportation. However, these advancements face significant challenges due to latency and bandwidth constraints imposed
		by traditional cloud-based machine learning (ML) frameworks. The need for innovative solutions is evident as cloud computing struggles with increased latency
		and network congestion. Previous attempts to offload parts of the ML pipeline to
		edge and cloud layers have yet to fully resolve these issues, often worsening system response times and network congestion due to the computational limitations of
		edge devices. In response to these challenges, this study introduces the InTec (Integrated Things-Edge Computing) framework, a groundbreaking innovation in IoT
		architecture. Unlike existing methods, InTec fully leverages the potential of a threetier architecture by strategically distributing ML tasks across the Things, Edge, and
		Cloud layers. This comprehensive approach enables real-time data processing at the
		point of data generation, significantly reducing latency, optimizing network traffic,
		and enhancing system reliability. InTec’s effectiveness is validated through empirical evaluation using the MHEALTH dataset for human motion detection in smart
		homes, demonstrating notable improvements in key metrics: an 81.56\% reduction
		in response time, a 10.92\% decrease in network traffic, a 9.82\% improvement in
		throughput, a 21.86\% reduction in edge energy consumption, and a 25.83\% reduction in cloud energy consumption. These advancements establish InTec as a new
		benchmark for scalable, responsive, and energy-efficient IoT applications, demonstrating its potential to revolutionize how the ML pipeline is integrated into Edge-AI
		(EI) systems.
	\end{abstract}
	

	\section{Introduction}
	\label{Introduction}
	In the digital transformation era, IoT drives innovation in home automation,
	healthcare, and transportation, promising enhanced efficiency and safety. The
	value of IoT applications lies in their ability to perform reliably and instantly,
	especially in critical areas such as healthcare and Industry applications, where
	timely and accurate decisions are crucial. In healthcare, real-time data processing
	enables prompt, life-saving interventions, while in industrial settings, edge computing supports autonomous decision-making and predictive maintenance \cite{c1}.
	However, significant challenges in data processing and system reliability arise
	due to the vast amounts of real-time data generated by numerous devices, leading
	to potential network overload and reduced responsiveness. Traditional cloud-centric models face limitations due to latency and centralized processing bottlenecks \cite{c6, c7, c8, c9, c10}. Edge computing offers a solution by decentralizing computational tasks.
	Yet, it faces challenges in deploying ML models for data analysis and inference
	due to resource constraints and network bandwidth limitations, resulting in suboptimal performance \cite{c11, c12, c13}.
	
	Studies \cite{c14, c15} are pivotal, laying the groundwork for future exploration. They
	have attempted to distribute a portion of the ML pipeline across the edge and
	cloud layers to reduce the processing load on the cloud layer while maintaining
	an acceptable level of inference accuracy. However, deploying ML models on the
	edge layer remains challenging despite these efforts due to resource constraints.
	In the approach detailed in \cite{c14}, the ML model’s architecture spans mobile
	devices and edge servers. This design is engineered to optimize data inference
	under varying network conditions by employing techniques such as segmentation and Early-Exiting, which adjust based on available bandwidth. Meanwhile,
	\cite{c15} introduces a method that minimizes the feature set sent to the cloud through
	the Autoencoder algorithm applied at the edge. This preprocessing step, which
	includes normalization, categorization, and dimensionality reduction, prepares
	data for subsequent cloud-based training and inference, assigning distinct roles
	to the Edge and cloud layers in the ML pipeline. In the paper \cite{c14}, the best-case
	delay reaches 400 ms, while in the article \cite{c15}, the system’s latency has increased
	because the data analysis task was placed on the cloud layer.
	
	However, these approaches suffer from several critical inefficiencies: (1)
	Deploying the learning model on cloud servers or distributing it between mobile
	devices and edge layers has often increased system latency during inference operations. (2) Reducing data dimensions at edge servers, followed by restoration at
	cloud servers, results in higher processing loads on both edge and cloud layers,
	exacerbating latency and energy consumption. (3) These studies typically distribute only portions of the ML pipeline—such as data preprocessing, training, or
	model deployment—across edge and cloud layers, failing to fully utilize the processing capabilities of the entire edge-cloud-things architecture.
	
	These inefficiencies in previous approaches stem from the edge layer’s comparatively lower processing capabilities and the ever-growing volume of IoTgenerated data. These challenges create bottlenecks, particularly as the number of devices and the volume of requests increase, directly impacting key performance metrics such as model learning accuracy, system latency, network traffic, throughput, and energy consumption. These gaps have long been recognized in
	the IoT domain, but it remains an open area of investigation. Our research seeks
	to address these persistent issues from a new perspective, offering a more comprehensive and efficient solution.
	
	The motivation for this research lies in the increasing demand for IoT devices
	to perform complex ML tasks locally, enabling more intelligent environments in
	healthcare, transportation, and home automation. As devices take on more MLdriven tasks, such as real-time data analysis, predictive maintenance, and decisionmaking, the need for efficient and distributed processing becomes critical. Addressing these challenges is essential for reducing latency, improving scalability, and
	enhancing the intelligence of devices, thereby creating a more interconnected and
	more intelligent world.
	
	Unlike existing solutions that only partially integrate ML models within the edgecloud paradigm, InTec introduces a novel distribution technique that deploys components of the ML pipeline across all three layers of the IoT architecture—Things,
	Edge, and Cloud. Traditional approaches typically split tasks between just the edge
	and cloud layers, often leaving IoT devices underutilized and creating centralized
	processing bottlenecks. In contrast, InTec’s architecture assigns specific, complementary roles to each layer, with IoT devices, edge servers, and cloud servers each
	contributing distinct parts of the processing pipeline. This configuration enables
	data to be processed and decisions to be made closer to where the data is generated.
	This holistic design enhances real-time processing, reduces latency, and ensures
	high system reliability without overloading the network. By deploying ML models
	directly onto IoT devices, InTec enables localized data analysis, reducing network
	traffic congestion and improving responsiveness. This makes InTec a uniquely scalable, fully distributed solution that meets the growing need for responsive, efficient
	IoT applications in data-heavy environments.
	
	By doing so, InTec addresses the existing limitations and ushers in a new era
	of IoT system performance, where real-time responsiveness and reliability are not
	aspirational goals but practical realities. This research’s novelty lies in its meticulous
	design that seamlessly integrates the strengths of edge computing with the pervasive
	nature of IoT devices, establishing a new benchmark for IoT system architecture and
	demonstrating a scalable solution for the ever-growing data processing demands in
	smart environments.
	
	In this study, we rigorously applied the innovative InTec framework across four
	distinct scenarios: cloud-based, edge-based, hybrid edge-cloud, and a comprehensive edge-cloud-things configuration. Our comparative analysis benchmarked
	against the methodologies presented in the articles \cite{c15, c16, c17} specifically targeted the
	Human Activity Recognition (HAR) challenge within a smart home context, utilizing the MHEALTH dataset \cite{c18}. Our contributions can be summarized in three
	key points: (1) Comprehensive ML Pipeline Distribution: We developed a novel
	framework that effectively distributes the ML pipeline across all layers of the IoT
	architecture—Things, Edge, and Cloud—enabling localized data processing and
	optimizing the use of computational resources. (2) Enhanced Performance in HAR Applications: Through rigorous testing on the MHEALTH dataset, we demonstrated
	significant improvements in latency, network traffic, throughput, and energy efficiency, particularly in Human Activity Recognition (HAR) tasks, where real-time
	processing is critical. (3) Application of CNN-LSTM for Time Series Data: We
	employed the CNN-LSTM (Convolutional Neural Network-Long Short-term Memory Networks) model within the InTec framework, proving its effectiveness in handling time series data in HAR scenarios and showcasing its adaptability to complex,
	real-world IoT applications.
	
	The structure of this article is meticulously organized to ensure ease of comprehension and seamless navigation. Section 2 provides foundational definitions, followed by a critical literature review in Sect. 3. Sections 4 and 5 introduce the proposed InTec framework and outline the evaluation methodology. Section 5.6 details
	the experimental design, while Sect. 5.7 synthesizes the findings. The article concludes with Sect. 6, summarizing the study’s core insights and contributions and
	outlining future research directions.
	
	\section{Background}
	\label{Background}
	\subsection{Deep learning and edge‑AI integration in IoT systems}
	Various algorithms, including rule-based and data-driven approaches, have been
	developed to address the complexity and heterogeneity of data generated by IoT
	devices. New IoT devices, like smart cameras, generate feature-rich data that traditional algorithms struggle to handle due to their complexity and volume, necessitating advanced analytical methods like deep learning \cite{c12}. With its multilayer
	abstraction, deep learning significantly impacts fields like speech and image recognition and object detection, making it useful for IoT environments where data complexity exceeds traditional processing capabilities \cite{c19}. The advent of EI leverages
	computational power at the network’s edge, facilitating local processing to reduce
	latency and conserve bandwidth \cite{c3, c11, c12, c20}. This local processing capability is further enhanced within the broader edge-cloud continuum by frameworks like \cite{c21},
	where computational tasks can be dynamically distributed across the edge and cloud
	environments. This combination of EI and the edge-cloud architecture transforms IoT devices into intelligent systems capable of immediate, informed actions, representing a significant shift in data processing and utilization. Moving	computational tasks closer to data generation and simultaneously enabling cloudbased analytics when necessary, these systems overcome centralized processing constraints and unlock new possibilities for automation and efficiency in real-time adaptive applications \cite{c12, c13}.
	
	\subsection{Machine Learning Pipeline}
	Integrating ML into operational environments, particularly within the IoT ecosystem, necessitates a structured approach to managing the lifecycle of ML models, leading to the development of ML pipelines. These pipelines automate and
	streamline training, deployment, and utilization processes, ensuring efficient
	execution in a sequential and automated workflow \cite{c23}. However, as depicted
	in Fig. 1, this workflow is iterative, ensuring stages are repeated to refine and
	employ the most accurate trained model. Utilizing ML pipelines in IoT settings
	offers benefits such as reusability, containerization, and parallel processing.
	These features enhance scalability and efficiency: reusability allows pipelines
	to be applied across different projects with minimal adjustments, containerization enables deployment on various platforms, and parallel processing distributes
	tasks across multiple resources, reducing analysis time. ML pipelines provide a
	structured framework for managing AI models, streamlining development and
	deployment, and ensuring efficient updates and maintenance, fostering continuous improvement and innovation in IoT systems \cite{c24}. Below is an elucidation of
	these stages:
	
	\textbf{1. Data Validation:} The foundation of any robust ML pipeline begins with data
	validation. This critical first step ensures the integrity and quality of data before
	it enters the training phase. By examining the statistical properties of the dataset, such as range, count, and distribution of classes, data validation aims to
	preemptively identify and rectify discrepancies that could compromise the learning model’s performance.
	
	\textbf{2. Data Preprocessing:} Following validation, data undergoes preprocessing to transform it into a format suitable for model training. This stage addresses the heterogeneity and complexity of IoT data, involving tasks such as normalization, feature
	engineering, and dimensionality reduction. Given IoT data’s diverse and often
	unstructured nature, preprocessing is vital for teasing out relevant features that
	can significantly impact the model’s accuracy and efficiency.
	
	\textbf{3. Model Training:} With the data prepared, the pipeline progresses to model training,
	where algorithms learn from the data. This iterative process adjusts the model’s
	parameters to minimize error and enhance its ability to make accurate predictions. In the context of IoT, where models often need to operate under resource constraints, the training phase also involves optimizing the model for performance
	and efficiency.
	
	\textbf{4. Model Analysis:} After training, the model undergoes a thorough evaluation to
	assess its performance. Metrics such as accuracy, precision, recall, and AUC
	provide insights into the model’s effectiveness in addressing the problem. This
	stage is crucial for identifying areas for improvement and ensuring that the model
	meets the requisite standards for deployment.
	
	\textbf{5. Model Deployment:}  The culmination of the ML pipeline is deploying the model
	for inference. In IoT systems, this often means integrating the model into edge
	devices or cloud platforms, which can process real-time data and provide actionable insights. This stage requires careful consideration of the deployment environment to optimize for latency, energy consumption, and scalability.
	
	\begin{figure}[t]
		\centering
		\includegraphics[width= 0.7\textwidth]{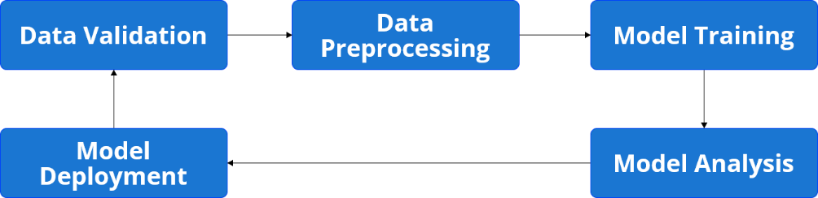}
		\caption{Machine Leaning Pipeline \cite{c23}}
	\end{figure}
	
	\section{Related Works}
	\label{Related Works}
	\subsection{ Evolution and trends in edge‑AI research }
	The domain of EI has experienced rapid expansion, which has been distinguished
	by a series of methodological and applied breakthroughs from 2018 to the present.
	A chronological overview, illustrated in Fig. 2, serves to articulate this progression:
	
	\textbf{2018-Foundational Integrations of Deep Learning and IoT:} The year marked the inception of integrating deep learning with IoT, as seminal works by Li et al. \cite{c25} and Zhao et al. \cite{c26} established the initial frameworks for embedding ML models within IoT infrastructures, laying the groundwork for subsequent EI innovations.
	
	\textbf{2019: The Advent of Deep Learning in Edge Computing:} In this phase, the focus transitioned to deploying deep learning models directly onto the Edge layer, with studies by Manogaran et al. \cite{c27} and Azar et al. \cite{c28} exploring architectural strategies to enhance computational offloading.
	
	\textbf{2020: Data Inference Convergence with the Edge:} This year witnessed an emphasis on data analytics at the edge, highlighted by contributions from Hu et al. \cite{c29}, and Li et al. \cite{c14}, advanced the practical implementation of ML models for real-time data inference within edge environments.
	
	\textbf{2021: Empowering Things with EI:} Researchers such as Kristiani et al. \cite{c31}, Ghosh et al. \cite{c9}, and Raj et al. \cite{c32} shifted focus towards empowering IoT devices ('Things') with edge intelligence, proposing architectures that leveraged the computational proximity of Edge layers.
	
	\textbf{2022: Setting up ML Pipelines on EI:} The setup of ML pipelines on EI became a focal point, with studies like Arunachalam et al. \cite{c33} exploring the benefits of distributed ML processes to optimize the entire data lifecycle on edge devices.
	
	\textbf{2023: Utilizing ML Pipelines for EI Challenges:} Investigations by Achar et al. \cite{c16} and Wazwaz et al. \cite{c17} probed deeper into the application of ML pipelines to solve complex EI problems, underscoring the transition from theory to application-centric solutions.
	
	\textbf{2024: Empowering Edge-AI by Inference at Things:} As of 2024, our research takes a
	step further by advancing inference at the ‘Things’ level, optimizing latency and
	reliability for EI applications, and pushing the evolution of EI research beyond
	prior works for real-world implementation.
	
	\begin{figure}[t]
		\centering
		\includegraphics[width= 0.8\textwidth]{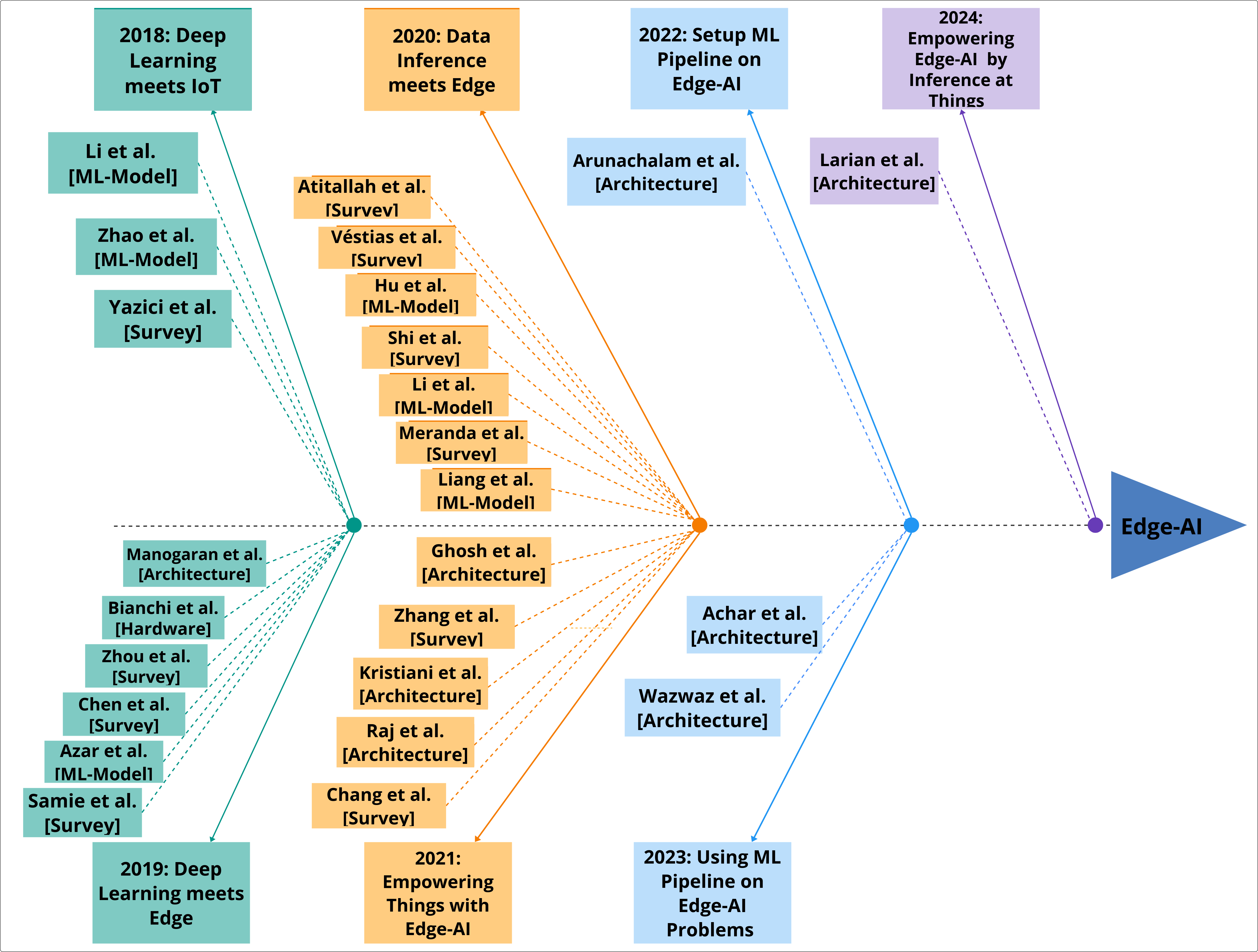}
		\caption{Edge-AI Studies Timeline}
	\end{figure}
	
	\subsection{ Overview of edge‑AI architectural methodologies}
	Various methods have been proposed for deploying ML models on the Edge layer, each
	tailored to specific contexts. In Fig. 3, each of these methods is generally categorized
	into sub-methods. This figure also represents the research domain addressed explicitly
	in this study. The research begins by searching in the field of EI studies. This field consists of two broad sub-areas known as training and inference, and our research focuses
	on designing hybrid architectures for EI systems by deploying an ML pipeline across
	the Things, Edge, and Cloud layers, optimizing latency and network traffic by leveraging the processing capabilities of these layers. Understanding EI architectural methodologies is crucial for appreciating the context and effectiveness of ML pipeline deployment in EI systems. Tables 1 and 2 provide a concise overview of critical studies in
	the EI domain from 2018 to 2024, highlighting architectural choices, ML models, use
	cases, datasets, methodologies, techniques, and performance metrics, including details
	on the InTec model. These comparisons cover cloud-edge to device-edge integrations
	using public and private datasets, primarily in healthcare and industry. Techniques such
	as model partitioning, compression, and optimization are detailed alongside parameters like accuracy, efficiency, latency, and energy consumption, showcasing each approach’s
	operational effectiveness and resource impact. The purpose of Table 2 is to present various performance metrics reported by each work, even if they differ across studies. For
	instance, some works, like \cite{c23}, reported Recall and FPR, while others, such as \cite{c27}, focused on Latency. It’s important to note that our study did not focus on optimizing
	or comparing ML model accuracy. Instead, our work emphasizes distributing the ML
	pipeline while maintaining better ML model accuracy. The diversity in metrics reflects
	how different studies approach performance evaluation.
	
	\begin{figure}[h]
		\centering
		\includegraphics[width= 0.7\textwidth]{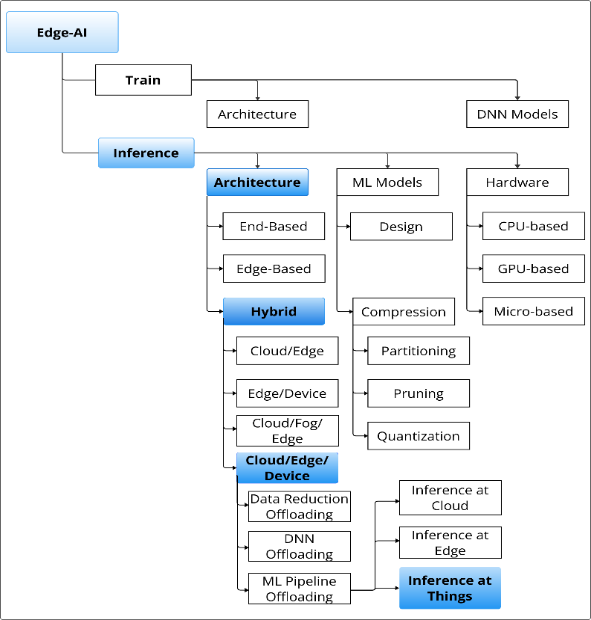}
		\caption{Pipeline Distribution in Edge-AI Mind Map}
	\end{figure}
	
	\subsubsection{End-Based and Edge-Based Edge-AI Architectures}
	End-based and edge-based architectures endeavor to optimize computational efficiency at the network's periphery. While end devices facilitate data processing at their origin to potentially reduce latency, their limited computational capacity restricts the complexity of deployable machine learning models. Conversely, edge-based architecture, by situating the inference process on edge servers, offers improved computational resources, enabling the deployment of more sophisticated, albeit lightweight, models closer to data sources. This approach, however, must navigate the constraints of edge resources. A notable example is a study \cite{c27} that introduced an architecture utilizing smart wristbands for transmitting medical data directly to an edge server. The Proposed InTec framework differentiates itself by
	distributing the ML pipeline across Things, Edge, and Cloud layers, enabling the
	deployment of more complex models while optimizing computational resources
	across the entire architecture.
	
	\subsubsection{Hybrid Edge-AI Architecture}
	Hybrid EI architectures represent a sophisticated approach to deploying ML pipelines by integrating multiple processing layers and optimizing computational workload distribution. This method excels in leveraging the combined processing power
	of cloud, edge, and end devices to facilitate efficient ML inference. Such architectures are categorized into distinct subgroups, each addressing specific operational
	needs and challenges:
	
	\textit{1. Cloud-Edge Architectures:} This subgroup melds the capabilities of cloud and
	edge computing, strategically allocating tasks to either layer based on their computational demands and the critical nature of the functions. The design challenge
	here revolves around the dynamic allocation of resources and tasks, aiming to
	minimize latency while ensuring efficient data processing. Cloud servers primarily manage the heavy lifting of data inference training and updating ML
	models, whereas edge servers focus on real-time data analysis. This distribution
	helps significantly reduce system response times. Notable studies, such as those
	cited in references \cite{c25} and \cite{c32}, explore innovative methods like Early-Exit and
	model partitioning to enhance responsiveness while minimizing latency despite
	the overhead introduced by data compression and the need for continuous model
	deployment and updating across the cloud-edge continuum. Our presented framework, InTec, differentiates itself by extending the distribution of the ML pipeline
	beyond just the cloud and edge layers, incorporating the Things layer as well. This
	approach enables more efficient real-time processing and reduces latency further
	by localizing more computational tasks and optimizing resource use across the
	entire IoT architecture.
	
		\textit{2. Edge-Device Architectures:} Tailored for time-sensitive applications, such as health monitoring and autonomous vehicles, this architecture emphasizes rapid inference by deploying pre-trained models directly onto edge servers and devices. The challenge lies in managing the resource limitations inherent to both layers, necessitating the use of models that can provide accurate inferences with minimal computational demand. Research \cite{c31} Demonstrates the application of techniques like model layer division and Early-Exit strategies to address these constraints, ensuring
		that data can be processed efficiently across the spectrum of available resources.
		The InTec framework, the main contribution of this research, further extends this
		concept by distributing the ML pipeline across edge devices and integrating the
		Things and Cloud layers.
	
	\begin{landscape}
		\begin{longtable}{@{}p{1.5cm}p{3.5cm}p{2.5cm}p{2.5cm}p{2.5cm}p{3cm}p{5cm}p{2.2cm}@{}}
			\caption{The comparison of existing studies on the Edge-AI research area}\\\\
			\toprule
			References & Architecture & ML Model & Use Case & Dataset & Method & Technique & Parameters \\
			\midrule
			\endfirsthead
			\caption[]{The comparison of existing studies on the Edge-AI research area (continued)}\\\\
			\toprule
			References & Architecture & ML Model & Use Case & Dataset & Method & Technique & Parameters \\
			\midrule
			\endhead
			\bottomrule
			\endfoot
			\cite{c25} & Cloud/Edge & CNN & Industry & Private Dataset & ML-Model & Partitioning + Early-Exit & Accuracy, Efficiency \\
			\cite{c26} & Edge & CNN & NR & Coco & ML-Model & Partitioning + Dynamic Clusters Offloading & Latency, Network Traffic, Throughput \\
			\cite{c28} & Cloud/Edge & FFNN & Healthcare & Stress Recognition in Automobile Drivers & ML-Model & Compression & Accuracy, Energy, Latency, Network Traffic \\
			\cite{c36} & Cloud/Edge & CNN & Healthcare & MHEALTH & Hybrid Architecture & Optimizing & Accuracy \\
			\cite{c27} & Edge & EC-BDLN & Healthcare & Private Dataset & Edge-based Architecture & Optimizing & Accuracy \\
			
			\cite{c14} & Device/Edge & AlexNet & NR & Cifar-10 & ML-Model & Partitioning + Early-Exit + DNN Offloading & Accuracy, Latency, Traffic, Throughput \\
			\cite{c29} & Cloud/Edge & AlexNet/VGGNet-19 & NR & NR & ML-Model & DNN Offloading & Latency \\
			\cite{c37} & Cloud/Edge & VGG-16 & Industry & T-Less & ML-Model & DNN Offloading & Accuracy, Latency \\
			\cite{c32} & Cloud/Edge & SVR & Air Quality & Private Dataset & Hybrid Architecture & ML Pipeline Offloading & Accuracy \\
			\cite{c31} & Cloud/Device/Edge & CNN & Object Detection & ImageNet & Hybrid Architecture & ML Pipeline Offloading & Energy \\
			\cite{c15}  (baseline article) & Cloud/Edge & Autoencoder + FFNN & Healthcare & MHEALTH & Hybrid Architecture & ML Pipeline Offloading & Accuracy, Latency, Traffic \\
			\cite{c33} & Edge & ResNet & Industry & ImageNet & Edge-based Architecture & ML Pipeline Offloading + Optimizing & NR \\
			\cite{c16} & Cloud & GoogleNet-BiLSTM & Security & HMDB51 & Architecture & ML Pipeline Offloading + Optimizing & Accuracy \\
			\cite{c17} & Cloud/Edge & Light GBM & Healthcare & HAR (DWS) & Architecture & ML Pipeline Offloading & Accuracy, Latency \\
			This research (InTec) & Cloud/Edge/Things & CNN-LSTM & Healthcare & MHEALTH & Hybrid Architecture & ML Pipeline Offloading & Accuracy, Latency, Traffic, Throughput, Energy \\
		\end{longtable}

		\begin{table}[ht]
			\centering
			\caption{Performance Metrics}
			\resizebox{\linewidth}{!}{
				\begin{tabular}{ccccccccccc}
					\toprule
					Paper & Architecture & ML Model & Accuracy \% & Recall \% & FPR \% & Precision \% & Latency (ms) & Traffic Reduction \% \\
					\midrule
					\cite{c25} & Cloud/Edge & CNN & NR & 100 & 30 & NR & NR & NR \\
					\cite{c26} & Edge & CNN & NR & NR & NR & NR & 25000 & 52 \\
					\cite{c28} & Cloud/Edge & FFNN & 98 & NR & NR & NR & 860 + Tl & 103 \\
					\cite{c36} & Cloud/Device & CNN & 92.5 & 92 & NR & 91.6 & NR & NR \\
					\cite{c27} & Edge & EC-BDLN & 96.5 & NR & NR & NR & NR & NR \\
					\cite{c14} & Device/Edge & AlexNet & 77 & NR & NR & NR & 400 & NR \\
					\cite{c29} & Cloud/Edge & AlexNet / VGGNet-19 & NR & NR & NR & NR & 90 + Tl & NR \\
					\cite{c37} & Cloud/Edge & VGG-16 & 95 & NR & NR & NR & 35 + Tl & NR \\
					\cite{c32} & Cloud/Edge & SVR & NR & NR & NR & NR & NR & NR \\
					\cite{c31} & Device/Edge & CNN & NR & NR & NR & NR & NR & NR \\
					\cite{c15} & Cloud/Edge & Autoencoder + FFNN & 99.89 & NR & NR & NR & NR & 48.84 \\
					\cite{c33} & Edge & ResNet & NR & NR & NR & NR & NR & NR \\
					\cite{c16} & Cloud & GoogleNet–BiLSTM & 74.79 & 68.70 & NR & 73.01 & NR & NR \\
					\cite{c17} & Cloud/Edge & Light GBM & 99.6 & 99.6 & NR & 99.6 & 38 & NR \\
					 \\
					\bottomrule
				\end{tabular}
			}
			\vspace{10pt} 
			
			\begin{tabular}{l l}
				\textbf{Legend:} & NR = Not Reported \\
			\end{tabular}
		\end{table}
		\vfill
	\end{landscape}

	\textit{3. Cloud-Edge-Device Architectures:} This architectural framework aims to fully utilize processing capabilities from the cloud to edge devices, targeting unparalleled efficiency and rapid data inference. By allocating machine learning tasks according to the unique computational strengths and requirements of each layer, it strives for a cohesive operation. A notable application, referenced in \cite{c31} demonstrates this approach by integrating IoT technologies like LoRaWAN and MQTT for real-time environmental
	monitoring and object detection, with critical processing performed at the edge
	to boost responsiveness. Despite its ambitious goal to enhance processing power
	and reduce latency, this strategy encounters inherent challenges: coordinating
	complex multi-layer operations, managing resources across diverse environments,
	addressing latency to ensure timely data processing, ensuring scalability amidst
	fluctuating IoT network demands, and optimizing energy consumption for sustainability. These hurdles underscore the complexity of achieving seamless integration and operational efficiency within such a distributed architecture. The proposed InTec framework simplifies and optimizes
	the coordination of these multi-layer operations by strategically distributing the
	ML pipeline across the Things, Edge, and Cloud layers. This approach enhances
	processing power, reduces latency, and improves scalability and energy efficiency,
	making it more adaptable to fluctuating IoT network demands and diverse operational environments.
	
	\subsection{Development of Machine Learning Models for Edge-AI} 
	The development of EI machine learning models focuses on optimizing deployment across diverse devices, ensuring accuracy and adaptability to resource constraints. This involves creating models for EI with an emphasis on resource limitations, as seen in Lyu et al. \cite{c38}, who designed a privacy-preserving deep learning model for fog computing. Model compression techniques, including partitioning, pruning, and quantization, are essential for fitting models into resource-constrained environments, enhancing deployability and efficiency, as detailed by Zhao et al. \cite{c26}. Advancements in EI hardware, such as multi-core CPUs, GPUs, and enhanced microcontrollers, further support the execution of complex machine learning tasks on edge devices, enabling advanced applications like wearable IoT devices for human activity recognition, as demonstrated by Bianchi et al. \cite{c36}.
	
	\section{Proposed Framework}
	\label{Proposed Framework}
	
	\subsection{Pipeline Design for Proposed Framework}
	
	The pipeline for our proposed framework is meticulously structured to cover all
	essential phases: data validation, preprocessing, model training, analysis, and
	deployment. As visualized in Fig. 4, we have assigned modules to each stage, carefully distributing these across the Cloud, Edge, and Things layers to leverage their
	unique computational capabilities.
	
	At the foundation, the Things layer deploys the ML model, allowing the Inference module to make data-driven decisions close to the data source, thus enhancing
	responsiveness and reducing latency. The edge layer initially processes data using
	the Outlier Detection and Data Reduction modules, which validate and preprocess
	the data. This step significantly reduces the volume of data transmitted to the Cloud,
	conserving bandwidth and accelerating the workflow. In the Cloud layer, the Model
	Trainer and Model Validator modules leverage extensive computational resources
	for training and analyzing the ML model, ensuring rigorous evaluation of performance and accuracy.
	
	This hierarchical, layered approach to pipeline design ensures that each phase
	of the ML process is optimally placed within the architecture, from the immediate
	handling of data at its point of generation to the sophisticated analysis and model
	refinement processes in the cloud. This strategic distribution not only maximizes the efficiency and efficacy of the ML pipeline but also aligns with the overarching goal
	of our framework to create a resilient, scalable, and intelligent EI ecosystem.
	
	\begin{figure}[h]
		\centering
		\includegraphics[width= 0.9\textwidth]{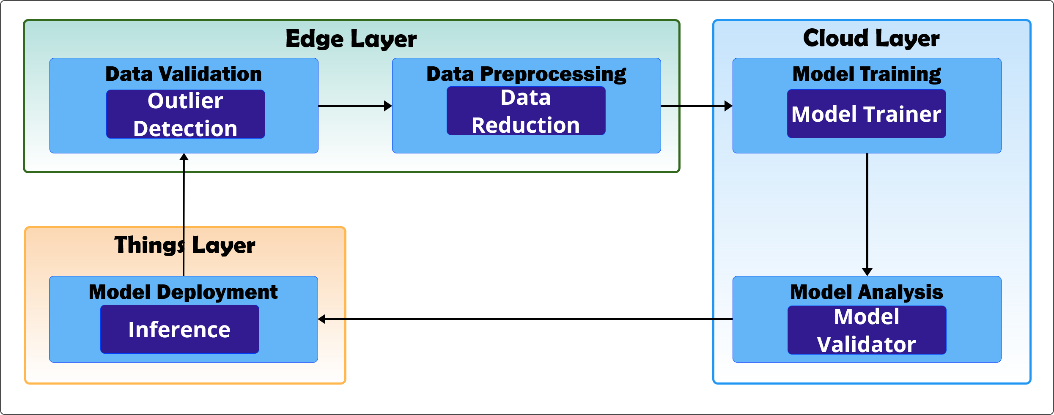}
		\caption{ML Model Pipeline Design for the Proposed Framework}
	\end{figure}
	
	\subsection{Inference at Things, in a Things-Edge-Cloud Architecture (InTec) Framework}
	This research introduces an advanced InTec framework designed to optimize the
	ML pipeline within a comprehensive cloud-edge-things ecosystem, as illustrated in
	Fig. 5. The architecture aims to streamline data processing, improve system response
	times, and minimize the amount of data transmitted to the cloud by strategically distributing tasks across three principal layers: cloud, edge, and Things. Each layer’s
	role and operations are detailed below.
	
	\begin{figure}[h]
		\centering
		\includegraphics[width= 0.7\textwidth]{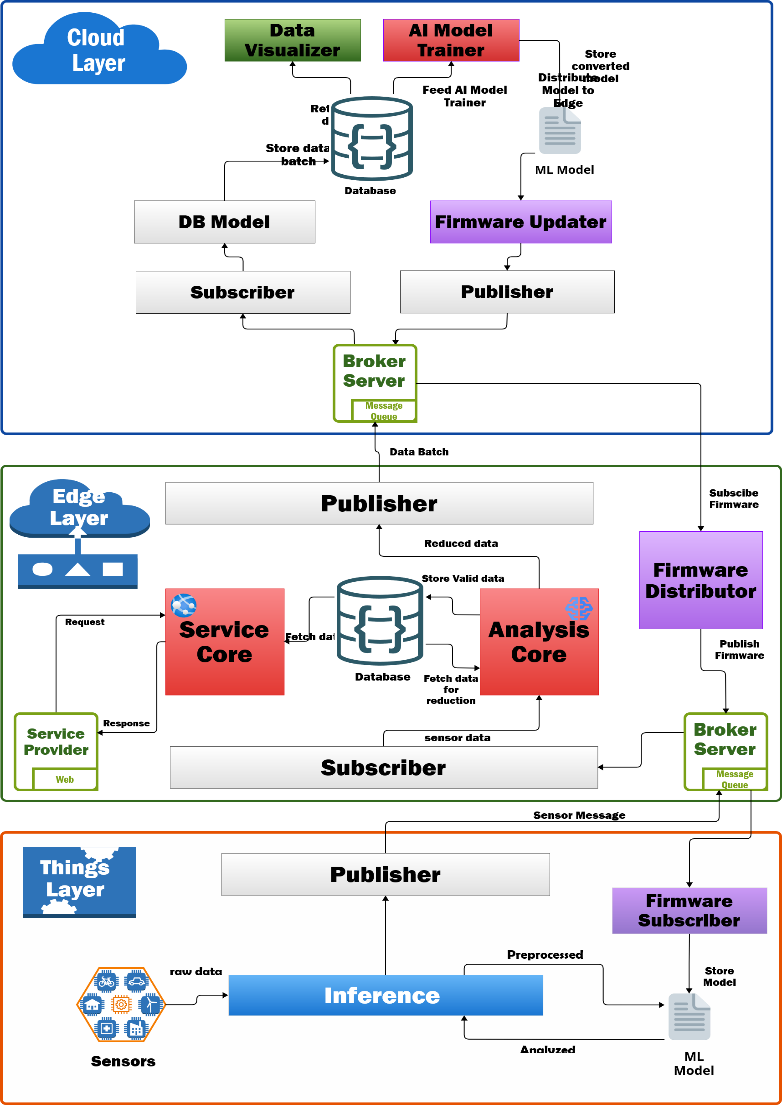}
		\caption{ InTec Architecture}
	\end{figure}
	
	\textit{Cloud Layer:} The cloud layer serves as the central hub for the training and analysis phases of the ML pipeline. It receives data from edge servers for periodic model training based on predefined policies, followed by result evaluation. The trained models are compressed to facilitate updates back to the edge layer.
	
	\textit{Edge Layer:} As the intermediary, the edge layer focuses on refining data collected
	from IoT devices. It employs data preprocessing and validation to structure data for
	cloud-based training. It includes (1) Data Dimensionality Reduction: Streamlines
	data, preserving user privacy and reducing cloud server load by minimizing data
	size. (2) Outlier Detection: Filters anomalies to maintain model integrity, ensuring
	only valid data influences model training. Additionally, this layer is responsible for
	disseminating the updated ML models to the sensors. 
	
	\textit{Things Layer: }	At the forefront of this framework, the Things layer employs IoT
	devices to execute data inference, utilizing pre-trained ML models for immediate
	analysis. This setup presumes IoT devices have adequate processing capabilities
	to run these optimized models. Data analysis is conducted directly on the devices,
	which then relay pertinent information to the edge server for subsequent actions.
	The operational blueprint for each IoT device encompasses raw data acquisition, indevice analysis, and communication with the edge server, demonstrating a self-sufficient approach to initial data processing. Through this layered architecture, the InTec
	framework achieves a balanced distribution of ML operations, from initial data capture to in-depth analysis, harnessing the unique strengths of cloud, edge, and Things layers to enhance overall system efficiency and responsiveness.
	
	\subsection{AI Model Trainer Module in the Cloud}
	The AI Model Trainer Module, as outlined in Fig. 6, is the core of the ML operations within the cloud layer of our InTec framework, performing critical functions,
	including data preparation, model training, validation, and compression. The workflow starts by retrieving data from the centralized database, which is then sent to
	the Preprocess sub-module for initial cleaning and transformation. After this, the
	cleaned data moves to the Model Trainer sub-module, where the core ML algorithm
	uses a specified ML configuration to build the model. Following successful training, the model is evaluated in the Model Validator sub-module to ensure it meets
	the required accuracy and credibility standards. The final step involves the Model Compactor sub-module, which compresses the validated model to significantly
	reduce its size for efficient deployment on resource-constrained IoT devices. From
	data preprocessing to model compression, this process is systematically detailed in
	Pseudocode 1, providing a step-by-step algorithmic representation of the module’s
	operations.
	
	\begin{figure}[t]
		\centering
		\includegraphics[width= 0.3\textheight]{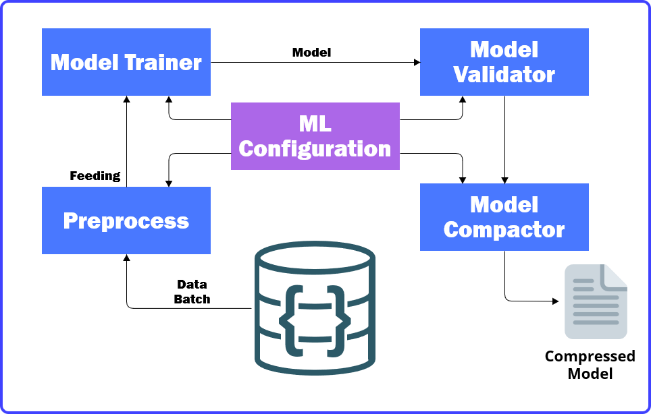}
		\caption{Sub-modules of AI Model Trainer}
	\end{figure}
	
	\begin{algorithm}[h]	
		\SetAlgoNlRelativeSize{-1}
		\SetKwInput{KwInput}{Input}
		\SetKwInput{KwOutput}{Output}
		\SetKwFunction{customereaddata}{custome\_read\_data}
		\SetKwFunction{splitsequences}{split\_sequences}
		\SetKwFunction{ModelCheckpoint}{ModelCheckpoint}
		\SetKwFunction{EarlyStopping}{EarlyStopping}
		\SetKwFunction{ModelConverter}{ModelConverter}
		
		\KwInput{Data}
		\KwOutput{Lite Model}
		\caption{AIModelTrainer(Data): LiteModel}
		
		$X\_train, y\_train, X\_test, y\_test$ $\leftarrow$ $\text{\customereaddata}(just\_load=True)$\;
		$train\_set, test\_set$ $\leftarrow$ $[X\_train\_pca, y\_train\_array]$, $[X\_test\_pca, y\_test\_array]$\;
		$X\_train\_seq, y\_train\_seq$ $\leftarrow$ $\text{\splitsequences}(train\_set, 25)$, $\text{\splitsequences}(test\_set, 50)$\;
		$n\_timesteps, n\_features, n\_outputs$ $\leftarrow$ $X\_train\_seq.\text{shape}[1]$, $X\_train\_seq.\text{shape}[2]$, $y\_train\_seq.\text{shape}[1]$\;
		
		$model \leftarrow \text{Sequential()}$\;
		\For{$layer$ \text{in} $[(32, 3), (64, 3), (32, 3)]$}{
			$model.\text{add}($\\
			\hspace*{2em}$\text{layers.Conv1D(filters=layer[0], kernel\_size=layer[1],}$\\
			\hspace*{2em}$\text{padding="same")})$\;
			$model.\text{add}(\text{layers.BatchNormalization()})$\;
			$model.\text{add}(\text{layers.ReLU()})$\;
			$model.\text{add}(\text{layers.MaxPool1D(2)})$\;
		}
		$model.\text{add}(\text{layers.LSTM(64)})$\;
		$model.\text{add}(\text{layers.Dense(units=128, activation='relu')})$\;
		$model.\text{add}(\text{layers.Dense(13, activation='softmax')})$\;
		
		$callbacks \leftarrow [\text{\ModelCheckpoint}("model\_25.h5", save\_best\_only=True, monitor="val\_loss"), \text{\EarlyStopping}(monitor="val\_loss", patience=10, verbose=1)]$\;
		$model.\text{compile}(optimizer='adam', metrics=['accuracy'], loss='categorical\_crossentropy')$\;
		$model\_history \leftarrow model.\text{fit}(X\_train\_seq, y\_train\_seq, epochs=50, validation\_data=(X\_test\_seq, y\_test\_seq), callbacks=callbacks)$\;
		
		$converter \leftarrow \text{\ModelConverter}(model)$\;
		$lite\_model \leftarrow converter.\text{convert()}$\;
	\end{algorithm}
	
	\subsection{Updating ML Models on IoT Devices}
	In the InTec framework, a tiered module system manages updating ML models on
	IoT devices. At the cloud layer, the Firmware Updater module creates firmware
	updates using the latest models by comparing device version data to determine if
	updates are needed. It then produces customized firmware packages and sends them to the edge layer when necessary. The Firmware Distributor module at the edge layer
	distributes these updates to IoT devices according to specific policies, optimizing
	network resources. The Firmware Subscriber module securely installs the updates
	on the IoT devices, enhancing analytical capabilities. This structured approach,
	detailed in Pseudocode 2, ensures consistency, reliability, and optimal functionality across the IoT network, maintaining cutting-edge performance in data-driven
	decision-making.
	
	\begin{algorithm}[H]
		\SetAlgoNlRelativeSize{-1}
		\SetKwInput{KwInput}{Input}
		\SetKwInput{KwOutput}{Output}
		\SetKwData{DeviceList}{DeviceList}
		\SetKwData{MLModelFile}{MLModelFile}
		\SetKwData{FirmwareFile}{FirmwareFile}
		
		\KwInput{Device List}
		\KwOutput{Firmware File}
		\caption{Firmware Updater(DeviceList): Firmware File}\label{alg:firmware-updater}
		
		\BlankLine
		\DeviceList $\leftarrow$ [{"DeviceID": \_id, "DeviceModel": \_model, "DeviceAddr": \_addr, "TopicUpdate": \_topic, "FirmwareVersion": \_Firmware\_ver, "MLModel\_Version": \_Model\_ver}]\;
		\MLModelFile $\leftarrow$ load("the Address of File")\;
		
		\BlankLine
		\For{device \text{in} \DeviceList}{
			\If{\MLModelFile.Version > device["MLModel\_Version"]}{
				\FirmwareFile $\leftarrow$ CreateFirmware(\MLModelFile, device["DeviceModel"])\;
				\If{Publish(\FirmwareFile, device["TopicUpdate"]) == True}{
					device["MLModel\_Version"], device["FirmwareVersion"] $\leftarrow$ \MLModelFile.Version, device["FirmwareVersion"] + 1\;
				}
			}
		}
	\end{algorithm}	
	
	\subsection{Analysis Core in Edge}
	The edge Layer’s Analysis Core in the InTec framework, illustrated in Fig. 7, is pivotal for refining raw data from IoT sensors. It starts with Outlier Detection, using an
	Outlier Model to filter out anomalies and ensure data integrity. The Data Reduction
	module then processes the validated data, which uses a Reduction Model to condense the dataset, retaining essential features. The Database (DB Model) is central
	to these operations, which stores and organizes data, facilitating its structured progression through the pipeline. Environment Variables dynamically adjust the detection and reduction modules in real-time, enhancing data fidelity and compactness for
	efficient EI analytics. We will further explain the Outlier Detection and Data Reduction modules.
	
	\begin{figure}[ht]
		\centering
		\includegraphics[width= 0.4\textheight]{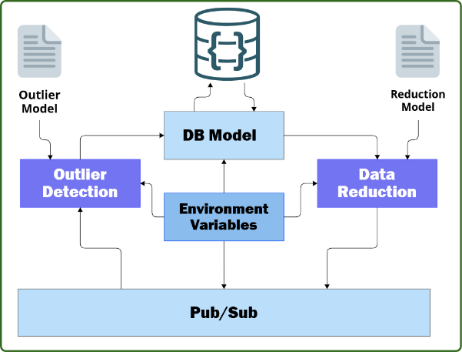}
		\caption{Analysis Core Workflow}
	\end{figure}
	
	\textit{1. Outlier Detection Module:} In the InTec framework’s Edge layer, Outlier Detection is critical for ensuring
	data quality. As shown in Fig. 8 and detailed in Pseudocode 3, raw data batches
	are converted to a structured data frame and assessed using a pre-trained Outlier
	Model to filter anomalies. This process employs a sliding window to evaluate
	data validity against a drop-rate threshold. Valid data is calculated by dividing
	the count of valid data points by the sliding window size, multiplied by 100, to
	obtain a threshold percentage. If this percentage exceeds the set threshold, the
	data is marked and stored for subsequent use, ensuring only high-quality data
	progresses through the learning pipeline, thereby maintaining the integrity and
	reliability of the ML models.
	
	\begin{figure}[th]
		\centering
		\includegraphics[width= 0.6\textheight]{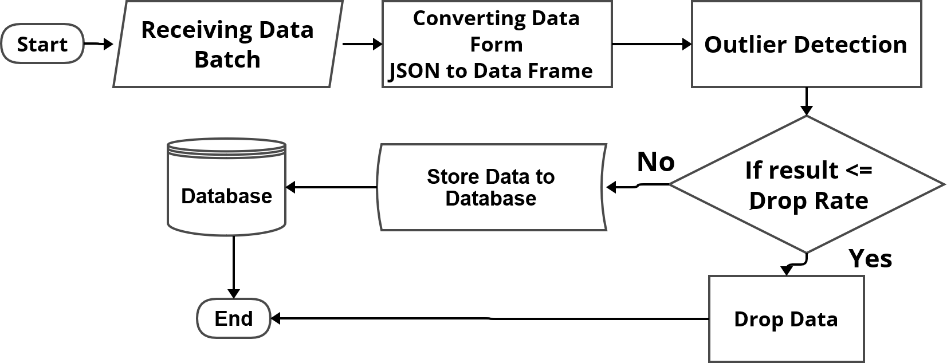}
		\caption{Outlier Detection Flowchart}
	\end{figure}
	
	\begin{algorithm}[h]
		\SetAlgoNlRelativeSize{-1}
		\SetKwInput{KwInput}{Input}
		\SetKwInput{KwOutput}{Output}
		\SetKwData{OutlierModel}{Outlier\_Model}
		\SetKwData{Data}{Data}
		\SetKwData{ConvertedData}{ConvertedData}
		\SetKwData{DataValidation}{DataValidation}
		\SetKwData{DataValidCount}{DataValidCount}
		\SetKwData{SlidingWindowSize}{sliding\_window\_size}
		\SetKwData{OutlierDropRate}{outlier-drop-rate}
		\SetKwData{OutlierModelName}{outlier\_model\_name}
		
		\KwInput{Data Batch}
		\KwOutput{Void}
		\caption{OutlierDetection(DataBatch): Void}\label{alg:outlier-detection}
		
		\BlankLine
		\OutlierModel $\leftarrow$ load(outlier\_file)\;
		\Data $\leftarrow$ getDatafromListener()\;
		\ConvertedData $\leftarrow$ DataFrame(Data['data']).T\;
		\DataValidation $\leftarrow$ DataFrame(\OutlierModel.predict(\ConvertedData)).T\;
		\DataValidCount $\leftarrow$ \DataValidation.values.tolist().count()\;
		
		\BlankLine
		\If{(\DataValidCount / \SlidingWindowSize) * 100 $\geq$ \OutlierDropRate}{
			\Data["validation"], \Data["outlier\_model"] $\leftarrow$ "Checked", \OutlierModelName\;
			dbmodel.insert(\Data)\;
		}
	\end{algorithm}
	
	\textit{2. Data Reduction Module:} The Data Reduction Module in the InTec framework
	streamlines the process of preparing and transmitting data to the cloud. As
	depicted in Fig. 9 and explained in Pseudocode 4, the module periodically fetches
	data from the database, compresses it to reduce size and complexity, and then
	sends this optimized data to the cloud, following a schedule to ensure efficient
	and regular updates. This systematic approach to data reduction not only maximizes the utility of the cloud layer’s expansive computational resources but also
	minimizes the latency and overhead associated with large-scale data handling,
	thereby enhancing the overall efficacy and responsiveness of the InTec ecosystem.
	
	\begin{figure}[h]
		\centering
		\includegraphics[width= 0.6\textheight]{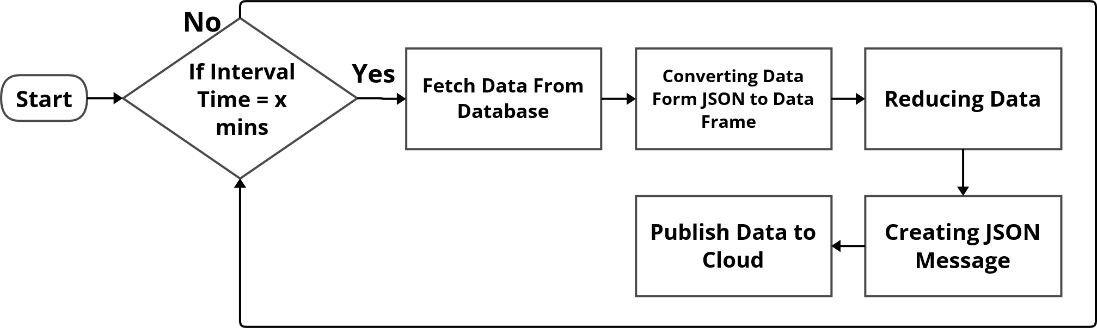}
		\caption{Data Reduction Flowchart}
	\end{figure}
	
	\begin{algorithm}[t]
		\SetAlgoNlRelativeSize{-1}
		\SetKwInput{KwInput}{Input}
		\SetKwInput{KwOutput}{Output}
		\SetKwData{ReductionModel}{Reduction\_Model}
		\SetKwData{Interval}{Interval}
		\SetKwData{ClientLoopFlag}{Client.Loop\_Flag}
		\SetKwData{Message}{Message}
		\SetKwData{EdgeClientId}{edge\_client\_id}
		\SetKwData{ReductionModelName}{reduction\_model}
		\SetKwData{SlidingWindowSize}{sliding\_window\_size}
		\SetKwData{Data}{Data}
		\SetKwData{DataBatch}{Data\_Batch}
		\SetKwData{ReducedData}{Reduced\_Data}
		\SetKwData{ReducedConvertedData}{Reduced\_Converted\_Data}
		\SetKwData{Topic}{Topic}
		\SetKwData{PublishMessage}{Publish\_Message}
		\SetKwData{PublishResult}{Publish\_Result}
		
		\KwInput{Data Batch}
		\KwOutput{Void}
		\caption{DataReduction(DataBatch): Void}\label{alg:data-reduction}
		
		\BlankLine
		\ReductionModel $\leftarrow$ load(Reduction\_File)\;
		\Interval, \ClientLoopFlag $\leftarrow$ ConstantNumberInMinutes, True\;
		\Message $\leftarrow$ {
			'edge\_id': \EdgeClientId,
			'reduction\_model': \ReductionModelName,
			'window\_size': \SlidingWindowSize,
			'date': str(datetime.datetime.utcnow()),
			'data': []
		}\;
		\While{\ClientLoopFlag AND \Interval = timeCounter()}{
			\DataBatch $\leftarrow$ dbmodel.fetch\_data\_batch(\Interval)\;
			
			\For{\Data \text{in} \DataBatch}{
				\ReducedData $\leftarrow$ DataFrame(\ReductionModel.transform(\Data["data"]))\;
				\ReducedConvertedData $\leftarrow$ loads(\ReducedData)\;
				\ReducedConvertedData["label"] $\leftarrow$ int(\Data["label"])\;
				\Message["data"].append(\ReducedConvertedData)\;
			}
			\PublishMessage $\leftarrow$ dumps(\Message)\;
			\PublishResult $\leftarrow$ Client.publish(\Topic, \PublishMessage)\;
			time.sleep(\Interval)\;
		}
	\end{algorithm}
	
	\subsection{Service Core in Edge}
	The Service Core is a pivotal component of the InTec framework, designed to
	facilitate user interaction through a REST-based architecture. It handles user
	requests and delivers services while adhering to RESTful API standards for
	standardized communication. The Web Framework is the operational backbone,
	efficiently managing requests and responses. The DB Engine Module orchestrates
	interactions between the Web Framework and the database, ensuring scalable and
	responsive service by managing multiple concurrent connections. The Services
	Module houses a variety of services, each tailored to execute specific functionalities, enhancing the framework’s robustness. Together, these modules create a
	dynamic and responsive Service Core, acting as the interface between users and
	the robust backend processes of the InTec ecosystem.
	
	\subsection{Inference Module in Things}
	The Inference module, within the Things layer of the InTec architecture, is crucial
	for real-time data analysis, as depicted in Pseudocode 5. It executes the ML model
	on sensor-acquired data, extracting actionable insights. The workflow includes three
	sub-modules:
	
	\textit{1. Preprocess Sub-module:} This submodule initiates the workflow by preparing
	the raw sensor data for analysis. It converts the data into a standardized format,
	normalizes it to ensure consistency, and categorizes it for better identification and
	subsequent processing.
	
	\textit{2. Feed Sub-module:}  Serves as the analytical engine, where the prepared data is
	inputted into the ML model. It processes the data and outputs the inference
	results, and the interpreted data points are ready for use in decision-making or
	further action.
	
	\textit{3. Data Tagging Sub-module:} This submodule functions post-analysis to attach relevant metadata or labels to the processed data, preparing it for communication. It also formats the analyzed data, ensuring it is ready for efficient transmission to other system components or direct utilization.
	
	This structure allows for a streamlined and efficient data handling process within
	the Things layer, ensuring that data is analyzed promptly and prepared for subsequent use or transfer in an optimized format.
	
	\begin{algorithm}[h]
		\SetAlgoNlRelativeSize{-1}
		\SetKwInput{KwInput}{Input}
		\SetKwInput{KwOutput}{Output}
		\SetKwData{ScalerModel}{Scaler\_Model}
		\SetKwData{Inferencer}{Inferencer}
		\SetKwData{SlidingWindow}{Sliding\_Window}
		\SetKwData{ListOfData}{List\_of\_Data}
		\SetKwData{ConstantNumber}{ConstantNumber}
		\SetKwData{InputData}{Input\_Data}
		\SetKwData{OutputData}{Output\_Data}
		\SetKwData{Message}{Message}
		\SetKwData{Topic}{Topic}
		\SetKwData{InputDetails}{input\_details}
		\SetKwData{OutputDetails}{output\_details}
		\SetKwData{RawData}{Raw\_Data}
		\SetKwData{ConvertedData}{Converted\_Data}
		\SetKwData{ScaleData}{Scale\_Data}
		
		\KwInput{Data}
		\KwOutput{Void}
		\caption{InferenceModule(Data): Void}
		
		\BlankLine
		\ScalerModel $\leftarrow$ load(Scaler\_File)\;
		\Inferencer $\leftarrow$ Interpreter(Model.tflite)\;
		\SlidingWindow, \ListOfData $\leftarrow$ \ConstantNumber, []\;
		
		\While{True}{
			\If{len(\ListOfData) = \SlidingWindow}{
				\InputData $\leftarrow$ array(\ListOfData).reshape(1, \SlidingWindow, 23)\;
				\Inferencer.set(\InputDetails, \InputData)\;
				\Inferencer.invoke()\;
				\OutputData $\leftarrow$ \Inferencer.get(\OutputDetails)\;
				\Message $\leftarrow$ load(\InputData, \OutputData, 23, \SlidingWindow)\;
				Client.publish(\Topic, dumps(\Message))\;
				\ListOfData $\leftarrow$ []\;
			}
			\Else{
				\RawData $\leftarrow$ getDataFromSensor()\;
				\ConvertedData $\leftarrow$ DataFrame(\RawData).T\;
				\ScaleData $\leftarrow$ \ScalerModel.transform(\ConvertedData)\;
				\ListOfData.append(\ScaleData)\;
			}
		}
	\end{algorithm}
	
	\subsection{General Dynamics of User-System Interaction within the InTec Framework}
	Figure 10 detail a generalized sequence of interactions between a user and the InTec framework, delineating the flow and processing of data. The sequence involves the following steps:
	
	\textit{1. Data Capture:} The process begins with the sensors collecting data from the environment. This data could encompass a wide array of information depending on the sensor type and the monitoring needs.
	
	\textit{2. Data Processing Request:} After data collection, the Edge Server receives a request to process this data. The request may come directly from the user or be part of a predefined workflow within the system.
	
	\textit{3. Edge Processing:} Upon receiving the data, the Edge Server undertakes the initial processing steps. These may include validation, preliminary analysis, and preparation for any subsequent deep processing that may be required.
	
	\textit{4. User Query:} In parallel, the user may issue a request for specific information or actions to be performed by the system. This request is communicated to the Edge Server.
	
	\textit{5. Edge Analysis and Response:} The Edge Server processes the user's request, potentially utilizing the pre-processed data. It then formulates a response based on the request's parameters and the results of any analysis conducted.
	
	\textit{6. Information Delivery:} Finally, the Edge Server sends the response back to the user. This response contains the information or analysis outcome that the user sought, completing the interaction loop.
	
	\begin{figure}[b]
		\centering
		\includegraphics[width= 0.5\textwidth ,scale=0.15]{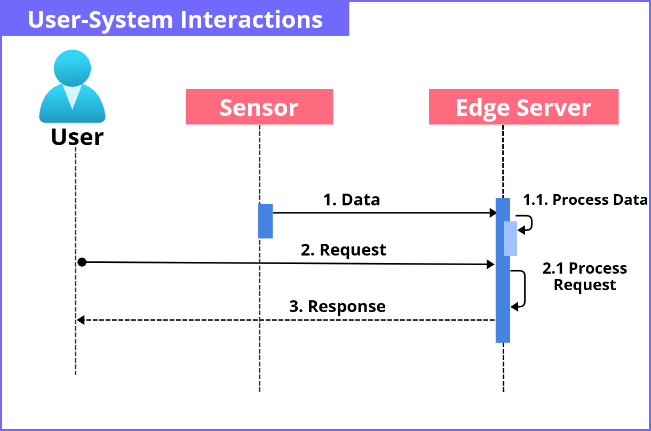}
		\caption{User-System Interaction Flow in the InTec Framework}
	\end{figure}
	
	\subsection{Optimizing Data Integrity in Sensor-Edge-Cloud Interaction}
	As depicted in the sequence diagram Fig. 11, this process ensures a seamless flow of
	data from the collection at the sensor level to utilization in the cloud while maintaining data integrity and privacy. The edge layer serves as a crucial intermediary, filtering and condensing data before it reaches the cloud, optimizing network resources,
	and safeguarding sensitive information. Here’s a description of this interaction:
	
	\textit{1. Data Collection:} Sensors actively gather data and initiate the interaction by publishing their findings to a designated topic. This information is typically in the form of a message that encapsulates the raw sensor data.
	
	\textit{2. Initial Processing at Edge:} Upon receiving the data, the Edge Server subscribes to the same topic to capture the sensor message. It employs an outlier detection algorithm on the received sensor data to determine its validity. If the data is flagged as an outlier and exceeds a pre-set drop rate, it is discarded to maintain data quality.
	
	\textit{4. Data Storage:} Valid sensor data, which passes the outlier test, is then stored in the Edge Database for further processing. This step is critical for building a repository of trusted data for subsequent stages.
	
	\textit{5. Periodic Data Reduction:} At regular intervals, determined by a time counter, the Edge Server fetches a batch of stored data from the database. It applies a dimensionality reduction process, condensing the data while preserving its salient features. This process aims to reduce the data's volume to streamline subsequent cloud transmission and to anonymize the data, thus enhancing user privacy.
	
	\textit{6. Data Transmission to Cloud:} The processed, reduced data is then published to a cloud topic, ready for Cloud Server subscription.
	
	\textit{7. Cloud Storage and Usage:} Upon receipt, the Cloud Server subscribes to the same topic to retrieve the message containing the reduced data. This data is then inserted into the cloud database, where it can be utilized for large-scale analytics, machine learning model updates, or provided to end-users through various applications.	
	
	\begin{figure}[t]
		\centering
		\includegraphics[width= 0.7\textwidth, scale=0.14]{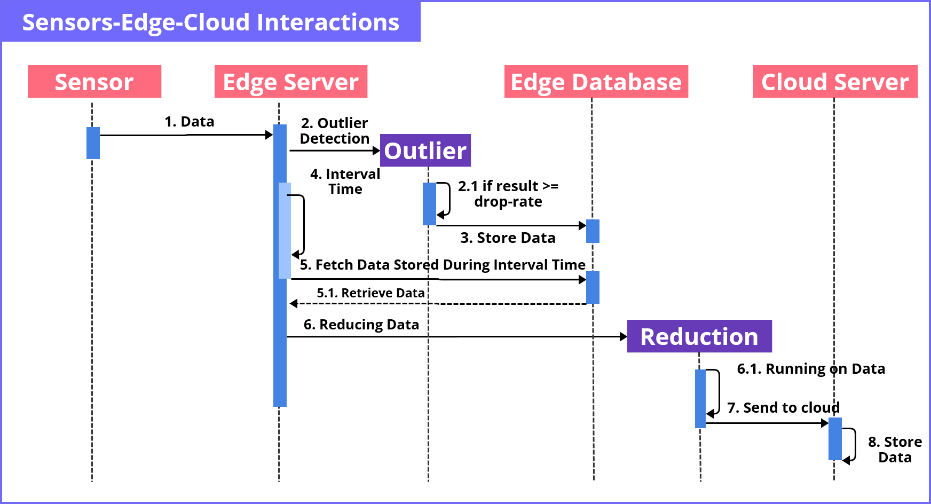}
		\caption{ Sequence diagram for sensor, edge, and cloud interaction}
	\end{figure}
	
	\subsection{Case Study}
	The previous sections provided a detailed explanation of the proposed framework.
	Now, in this section, a practical example is presented to demonstrate the performance of this framework in smart edge environments.
	
	A pre-trained and compressed model is deployed in the Objects layer and on each
	sensor. This model combines CNN and LSTM, trained on the cloud layer, and converted into the TFLight format. The data preprocessing stages, outlier detection, and
	dimensionality reduction are also deployed on the edge server.
	
	This use case scenario consists of an edge server and ten IoT sensors, simulated
	on Raspberry Pi devices. These ten sensors are implemented as separate containers
	on Docker. Each of these sensors analyzes its data with a window size of 25 and
	then sends it to the edge server at 50 Hz (matching the dataset’s sampling rate) with
	a time interval of 0.5 s (0.02 ×25).
	
	Figure 12 illustrates a human motion detection system scenario using the proposed InTec framework. Each sensor is responsible for collecting the movements
	of the targeted individual. These data are then categorized and analyzed with 21
	features and a window size of 25, and after labeling, they are sent to the edge layer.
	
	Upon entering the edge server, the data is first sent to the Outlier Detection module to identify data outside the valid range. Valid data is labeled and stored in the
	database. The stored data is periodically retrieved from the database and sent to the
	Data Reduction module. For example, data saved within the last 15 min is fetched
	from the database and sent to the Data Reduction module. In the Data Reduction
	module, data is batched in groups of 25, and dimensionality reduction is applied,
	reducing the number of features. In this example, data with 21 features is reduced
	using PCA with a 66\% reduction rate, resulting in data with seven features. Finally,
	the reduced data is categorized and sent to the cloud server.
	
	Upon entering the cloud server, the data is immediately stored in the database
	and, after a specified period, is retrieved from the database and provided to the AI
	Model Trainer module. In this module, the data is fed into a pre-trained model for
	training. The trained model is then compressed in TFLight format for distribution
	to the Objects layer. Additionally, the Visualizer module retrieves the data from the
	database for visualization as a graphical output.
	
	The user must send a request to the edge server through the web to obtain the
	data results. The Service Core handles the user’s request, and after retrieving the
	results from the database, a response is provided to the user. The figure below illustrates all these steps in order.
	
	\begin{figure}[th]
		\centering
		\includegraphics[width= 0.8\textwidth]{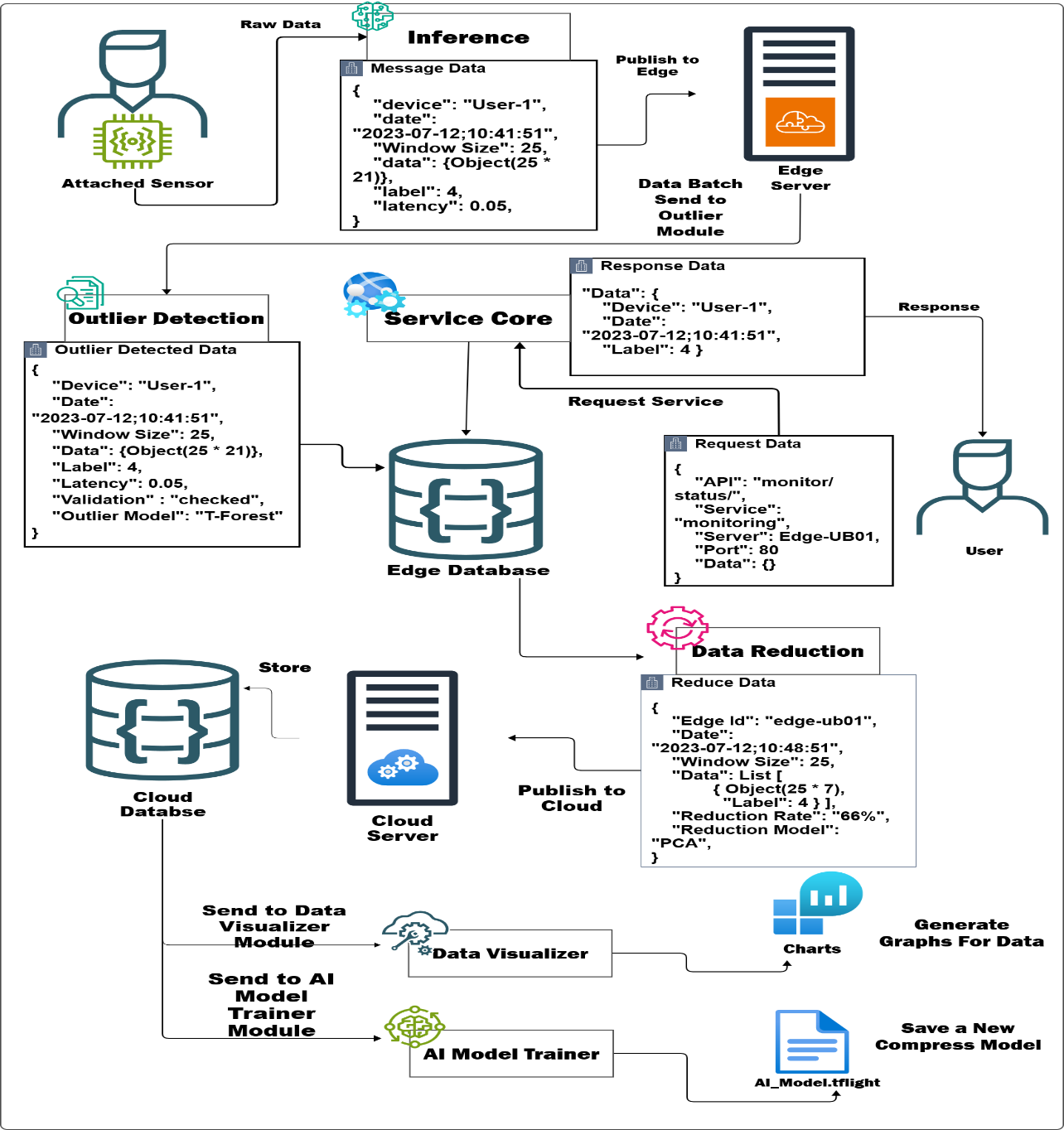}
		\caption{InTec Case Study Diagram}
	\end{figure}
	
	\section{Evaluation Methodology}
	To comprehensively evaluate the InTec framework, we tested its performance and applicability in a semi-real world environment through a meticulously crafted emulation scenario. This approach closely replicates real-world conditions, providing a robust platform to understand the framework's operation beyond theoretical settings. Through this emulation, we conducted a series of experiments to test the framework's capabilities and compare its results with benchmarks from related studies \cite{c15, c16, c17}, highlighting its impact on essential research variables and offering insights into its efficacy and areas for refinement. The source code for the InTec framework implementation is publicly available at \cite{c39}.
	
	\subsection{Emulation Scenario: HAR Problem}
	To faithfully replicate the baseline study conditions cited in \cite{c15}, this research adopts human motion detection as the focal emulation scenario. In alignment with the baseline study's methods, this emulation is executed on virtual machines using the MHEALTH dataset \cite{c18}, the same dataset employed by the baseline study for both training and validating the efficacy of the learning model. The MHEALTH dataset serves as the foundation for HAR research, comprising motion data from 10 distinct subjects captured by seven sensors per individual at a sampling rate of 50 Hz. Each recorded data instance encompasses 21 distinct features—derived from the 3-axes measurements of the seven sensors—and is classified into one of 12 movement classes, such as walking, standing, and running, offering a comprehensive range of human activities for analysis \cite{c40}. Table 3 illustrates the dataset’s structure, while Table 4 delineates the various movement categories recognized within it.
	
	\begin{table}[h]
		\centering
		\caption{The data features in the MHEALTH dataset}
		\adjustbox{width=0.8\textwidth}{
			\begin{tabular}{llll}
				\toprule
				\textbf{Row} & \textbf{Feature name} & \textbf{Unit} & \textbf{Description} \\
				\midrule
				1 & AC  & m/s\(^2\) & Acceleration from the chest \\
				2 & ALA & m/s\(^2\) & Acceleration from the left ankle \\
				3 & GLA & deg/s     & Gyro from the left ankle \\
				4 & MLA & gauss     & Magnetometer from the left ankle \\
				5 & ARA & m/s\(^2\) & Acceleration from the right lower \\
				6 & GRA & deg/s     & Gyro from the right lower arm \\
				7 & MRA & gauss     & Magnetometer from the right lower arm \\
				\bottomrule
			\end{tabular}
		}
	\end{table}
	
	\begin{table}[h]
		\centering
		\caption{Activity classes of MHEALTH dataset}
		\adjustbox{width=0.35\textwidth}{
			\begin{tabular}{ll}
				\toprule
				\textbf{Activity class} & \textbf{Label} \\
				\midrule
				Standing still              & L1  \\
				Sitting and relaxing        & L2  \\
				Lying down                  & L3  \\
				Walking                     & L4  \\
				Climbing stairs             & L5  \\
				Waist bends forward         & L6  \\
				Frontal elevation of arms   & L7  \\
				Knees bending (crouching)   & L8  \\
				Cycling                     & L9  \\
				Jogging                     & L10 \\
				Running                     & L11 \\
				Jump front and back         & L12 \\
				\bottomrule
			\end{tabular}
		}
	\end{table}
	
	\subsection{Emulation Setup and Implementation}
	The emulation of the InTec framework involved a detailed setup across IoT devices,
	edge, and cloud infrastructures using specialized tools and technologies. Raspberry
	Pi devices emulated sensors and used TensorFlow Lite Runtime and Docker containers for scalability. Virtual machines running Ubuntu OS were employed in both
	edge and cloud infrastructures. TensorFlow managed ML tasks in the cloud, while
	Scikit-Learn handled preprocessing functions at the edge. Node.js and Express
	framework facilitated RESTful API interactions on the edge server, while Docker
	containers ensured isolated service operations, as depicted in Fig. 13. A network
	of 100 sensors was emulated using Docker containers on virtual machines, demonstrating scalability and flexibility. All devices were interconnected via a TCP/IP
	network with 100 Mbps bandwidth, providing robust communication for accurate testing, with hardware specifications in Table 5. This setup highlighted the potential
	of ARM-based hardware for real-time analytics and validated data processing capabilities across different architectural layers. The edge server was configured for data management, inference (for Edge-based Inference at Edge framework), outlier detection, and network communications. In contrast, the cloud server was set up for data
	processing, model training using CNN and LSTM models, and advanced analytics,
	ensuring a robust evaluation of the InTec framework’s performance.
	
	\begin{figure}[h]
		\centering
		\includegraphics[width=0.65\textwidth]{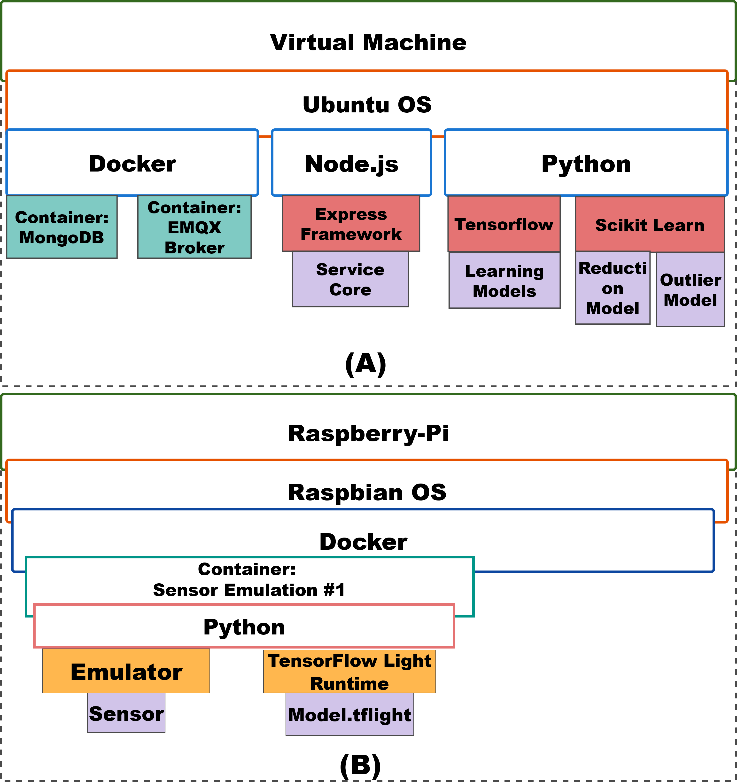}
		\caption{A) Edge and Cloud Emulation Environment, B) Sensors Emulation Environment}
	\end{figure}
	
	\begin{table}[ht]
		\centering
		\caption{Emulation Hardware Specifications}
		\adjustbox{width=0.99\textwidth}{
			\begin{tabular}{llllllll}
				\toprule
				Layer & Device Model & Num & OS & Processor Model & Cores & CPU Speed (GHz) & RAM (GB) \\
				\midrule
				Things & Raspberry Pi 2 B & 1 & Raspbian & ARM Cortex-A7 & 4 & 0.9 & 1 \\
				Things & Raspberry Pi 3 B & 2 & Raspbian & ARM Cortex-A53 & 4 & 1.2 & 1 \\
				Things & Virtual Machine & 1 & Ubuntu & Intel 6700 K & 1 & 4 & 4 \\
				Edge & Virtual Machine & 3 & Ubuntu & Intel 6700 K & 2 & 4 & 8 \\
				Cloud & Virtual Machine & 1 & Ubuntu & Intel 6700 K & 4 & 4 & 16 \\
				\bottomrule
			\end{tabular}
		}
	\end{table}

	\subsection{Analyzing Training Results of the CNN-LSTM Model for InTec Framework}
	Our research employs a combined CNN and LSTM model tailored for time-series
	data analysis within the InTec framework. The architecture depicted in Fig. 14
	addresses the nuanced demands of our analysis. We use z-score normalization to
	preprocess the MHEALTH dataset (a supervised dataset), standardizing it to have a
	mean of zero and a standard deviation of one. This study used 80\% of the preprocessed data for training and 20\% for testing.
	
	The model architecture consists of the following layers: Layer 1 employs a onedimensional convolution with 32 filters of size 3× 3, followed by Batch Normalization and the ReLU activation function. Layer 2 features a one-dimensional convolution with 64 filters of size 3 ×3, followed by Batch Normalization and ReLU. Layer
	3 utilizes one-dimensional MaxPooling. Layer 4 includes a one-dimensional convolution with 32 filters of size 3 ×3, followed by Batch Normalization and ReLU.
	Layer 5 applies one-dimensional MaxPooling. Layer 6 is an LSTM layer with 64
	units. Layer 7 is a fully connected layer with 128 neurons and the ReLU activation
	function. Finally, Layer 8 is fully connected with 13 neurons and the Softmax activation function.
	
	The CNN-LSTM model leverages the strengths of CNN and LSTM layers to handle time-series data efficiently and effectively. With their lower parameter count,
	CNN layers enable practical training with less data. In contrast, LSTM layers capture the temporal dynamics inherent in time-series data, preserving crucial relationships with a 64-dimensional output vector for each unit. Model performance is
	evaluated using metrics like Accuracy, Precision, Recall, and F1-Score across 50
	epochs, with early termination if accuracy does not improve over ten consecutive
	epochs. The Adam optimization algorithm and categorical Cross-Entropy loss function are used for parameter fine-tuning. Table 6 showcases comprehensive training
	outcomes under various configurations, highlighting the model’s adaptability and
	efficiency.
	
	\begin{figure}[t]
		\centering
		\includegraphics[width=0.65\textwidth]{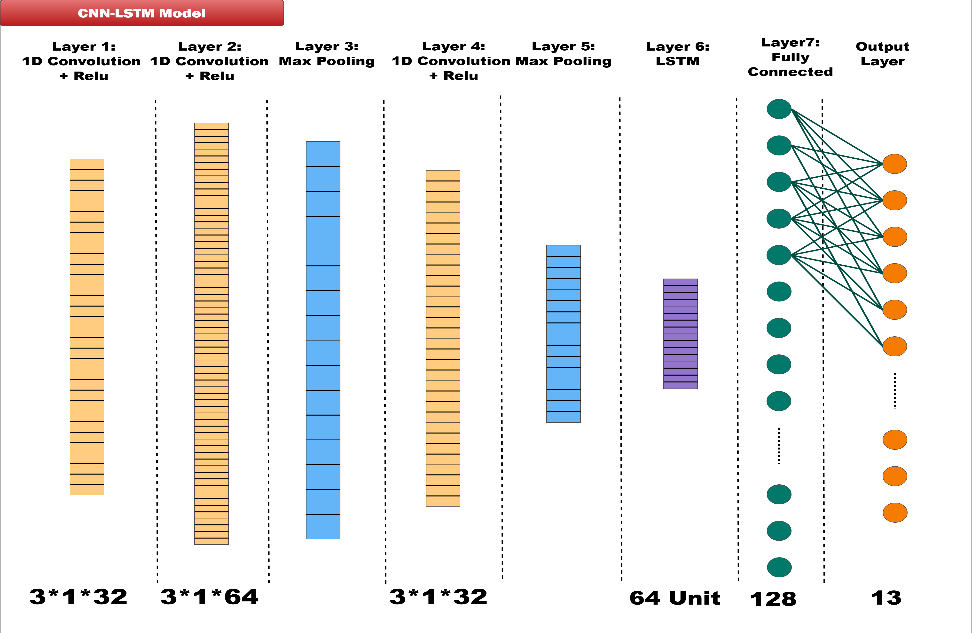}
		\caption{CNN-LSTM Architecture}
	\end{figure}
	
	\begin{table}[h]
		\centering
		\caption{Training Results of the CNN-LSTM Model}
		\adjustbox{width=\textwidth}{
			\begin{tabular}{cccccccccc}
				\toprule
				Row & Reduction Model & Feature Number (Reduction Ratio) & Window size & Iteration & ACC/val & Recall & Pres & F1 \\
				\midrule
				1 & PCA & 16 (24\%) & 25 & 50 & 99.9 & 99.9 & 99.9 & 99.9 \\
				2 & PCA & 16 (24\%) & 50 & 50 & 99.93 & 99.94 & 99.93 & 99.93 \\
				3 & PCA & 16 (24\%) & 100 & 50 & 99.88 & 99.87 & 99.88 & 99.88 \\
				4 & AE & 16 (24\%) & 100 & 50 & 95.45 & 95.45 & 95.45 & 95.45 \\
				5 & AE & 16 (24\%) & 50 & 50 & 95.25 & 95.25 & 95.25 & 95.25 \\
				6 & AE & 16 (24\%) & 25 & 50 & 93.91 & 93.79 & 94.03 & 93.9 \\
				7 & PCA & 7 (66\%) & 25 & 50 & 99.96 & 99.96 & 99.96 & 99.96 \\
				8 & PCA & 7 (66\%) & 50 & 50 & 99.89 & 99.89 & 99.9 & 99.89 \\
				9 & PCA & 7 (66\%) & 100 & 50 & 99.93 & 99.93 & 99.93 & 99.93 \\
				10 & AE & 7 (66\%) & 100 & 50 & 95.49 & 95.49 & 95.49 & 95.49 \\
				11 & AE & 7 (66\%) & 50 & 50 & 95.3 & 95.3 & 95.3 & 95.3 \\
				12 & AE & 7 (66\%) & 25 & 50 & 93.44 & 93.37 & 93.53 & 93.46 \\
				\bottomrule
			\end{tabular}
		}
	\end{table}
	
	\subsection{Evaluated Frameworks}
	In our study, the InTec framework was rigorously evaluated against existing frameworks highlighted in references \cite{c15, c16, c17}. This comparative analysis focused on how each framework implements the ML pipeline across different IoT architectures. To ensure a standardized comparison, all frameworks, including InTec, used the entire ML pipeline similarly and were emulated on identical infrastructures. This consistent setup allowed us to accurately measure performance, operational efficiencies, and scalability under real-world conditions. Table 7 outlines the distribution of the ML pipeline across the "Things," "Edge," and "Cloud" layers for all tested frameworks, highlighting the unique structure and flow of InTec.
	
	\textit{1. Cloud-based Environment:} In this setup, the IoT sensors are relegated to data collection and publishing duties, while the complete ML pipeline—comprising data validation, reduction, model training, validation, and deployment, including inference tasks—is centralized within the Cloud layer, bypassing the edge.
	
	\textit{2. Edge-based Inference in Cloud:} Here, the sensors collect and publish data, the Edge layer handles data validation and reduction, and the Cloud layer takes on model training, validation, and the deployment of models for inference.
	
	\textit{3. Edge-based Inference at Edge:} The sensors in this framework perform data collection and publication. The Edge layer has an expanded role, accommodating model deployment and data inference, besides validation and reduction, leaving the Cloud layer responsible for model training and validation only.
	
	\textit{4. InTec:} Our proposed framework advances a more distributed approach, where sensors not only collect and publish data but also deploy the model and conduct inference. The Edge layer manages data validation and reduction, while the Cloud layer focuses on model training and validation.

	\begin{landscape}
		\begin{table}[t]
			\centering
			\caption{Pipeline Distributions in Evaluated Frameworks}
			
			\begin{tabular}{lllll}
				\toprule
				\multirow{2}{*}{Row} & \multirow{2}{*}{Framework} & \multicolumn{3}{c}{ML Pipeline Distribution on Layers} \\
				\cmidrule(lr){3-5} 
				& & \multicolumn{1}{c}{Things} & \multicolumn{1}{c}{Edge} & \multicolumn{1}{c}{Cloud} \\
				\midrule
				1 & Cloud-based & 1. Collecting and Publishing data & Without Edge layer & 
				\begin{tabular}[c]{@{}l@{}}
					1. Validation \\
					2. Reduction \\
					3. Train Model \\
					4. Validating Model \\
					5. Deploy Model and Inference Data
				\end{tabular} \\
				\addlinespace
				2 & Edge-based Inference in Cloud & 1. Collecting and Publishing data & 
				\begin{tabular}[c]{@{}l@{}}
					1. Validation \\
					2. Reduction
				\end{tabular} & 
				\begin{tabular}[c]{@{}l@{}}
					1. Train Model \\
					2. Validating Model \\
					3. Deploy Model and Inference Data
				\end{tabular} \\
				\addlinespace
				3 & Edge-based Inference at Edge & 1. Collecting and Publishing data & 
				\begin{tabular}[c]{@{}l@{}}
					1. Validation \\
					2. Reduction \\
					3. Deploy Model and Inference Data
				\end{tabular} & 
				\begin{tabular}[c]{@{}l@{}}
					1. Train Model \\
					2. Validating Model
				\end{tabular} \\
				\addlinespace
				4 & InTec & 
				\begin{tabular}[c]{@{}l@{}}
					1. Collecting and Publishing data \\
					2. Deploy Model and Inference Data
				\end{tabular} & 
				\begin{tabular}[c]{@{}l@{}}
					1. Validation \\
					2. Reduction
				\end{tabular} & 
				\begin{tabular}[c]{@{}l@{}}
					1. Train Model \\
					2. Validating Model
				\end{tabular} \\
				\bottomrule
			\end{tabular}
			
		\end{table}
	\end{landscape}
	
	\subsection{Parameters and variables}
	\subsubsection{Independent Variables}
	(1) Sensor Quantity: This reflects the system’s total number of active sensors. A
	larger quantity implies more frequent data communication and higher network utilization. (2) User Count: Denotes the volume of concurrent users interacting with
	the system. Increasing user numbers can intensify the demand for system resources
	due to increased incoming queries. (3) Data Batching Window: The window size
	determines the dataset segment a sensor transmits in a single operation. A larger
	window reduces the frequency of transmissions but increases the data volume per
	batch. 4) Feature Reduction Ratio: This specifies the degree to which original sensor
	data dimensions are compressed before transmission. While a higher reduction can
	decrease data size and ease network load, it may inversely affect the precision of the
	data analysis.
	\subsubsection{Dependent Variables}
	\textit{1. Network Traffic:} Quantifies the volume of data traffic flowing through the network, typically presented in megabytes, highlighting system efficiency in data
	handling.
	
	\textit{Latency:} Measured in milliseconds, represents the average delay between a userinitiated request and the system’s response. This metric is crucial for evaluating
	the system’s responsiveness, indicating how quickly it can process and provide
	feedback or results to the user’s queries.
	
	\textit{Network Throughput:} Defines the capacity of the network to handle data over a
	specific period, with typical measurements in megabits per second, indicating
	the robustness of data transmission capabilities.

	\textit{Power Consumption:} Monitors the power draw of the deployed applications
	across devices, a critical factor for resource optimization and sustainability,
	measured in milliwatts.
	
	\section{Experimental Design}
	The experimental framework aims to systematically address five critical questions,
	as detailed in Table 8, each corresponding to distinct aspects of the InTec framework’s performance under varying conditions. These queries delve into the influence of data reduction rates, window sizes, the number of sensors, user requests, and the overall framework performance under high-load scenarios.
	
	Each experimental scenario is crafted precisely, focusing on specific variables
	such as network traffic, latency, throughput, and power consumption to comprehensively evaluate the framework’s robustness and efficiency. The experimental design
	utilizes two-dimensional reduction techniques, PCA (Principal Component Analysis) and AE (AutoEncoder), across all scenarios to elucidate their effects on the performance metrics.
	
	Select variables were held constant to guarantee consistency and equitable
	comparison in all experimental evaluations. The architecture employs a CNNLSTM model for ML tasks \cite{c32} and utilizes the Isolation Forest algorithm for
	outlier detection. Data for these experiments were sourced from the MHEALTH
	dataset \cite{c15}. Sensors operated at a uniform sampling rate of 50 Hz \cite{c18}, and the
	threshold for outlier removal was below 80\%. Additionally, each experiment was
	conducted three times to ensure reliability, with outcomes averaged to mitigate
	anomalies and provide a robust performance analysis under controlled variables.
	
	We’ve incorporated a strategic approach in our experiments to address the
	latency inherent in cloud-based user response frameworks and maximize cloud
	processor load. Each experiment utilized the entire MHEALTH dataset to train
	the ML model, ensuring that the cloud processor was under maximum operational stress. Additionally, to account for typical network delays experienced in
	real-world cloud interactions, a constant latency value representing the average
	ping from established cloud services was added to the response times measured
	in our experiments. This adjustment is reflected in Table 8.
	
	\begin{table}[t]
		\centering
		\caption{Latency of Well-known Cloud Providers}
		\adjustbox{width=0.9\textwidth}{
			\begin{tabular}{llll}
				\toprule
				Provider & Country & Server Address & Ping (ms) \\
				\midrule
				Google Cloud & Germany & \url{https://europe-west3-5tkroniexa-ey.a.run.app} & 69 \\
				Amazon Cloud & Germany & \url{http://dynamodb.eu-central-1.amazonaws.com} & 190 \\
				Oracle Cloud & Germany & \url{http://objectstorage.eu-frankfurt-1.oraclecloud.com/ping} & 142 \\
				Alibaba Cloud & Germany & \url{http://oss-eu-central-1.aliyuncs.com} & 123 \\
				Microsoft Azure & Germany & \url{http://speedtestden.blob.core.windows.net} & 203 \\
				Avg & & & 145 \\
				\bottomrule
			\end{tabular}
		}
	\end{table}
	
	\subsection{Experimental Scenarios}
	Experimental scenarios aim to cover a comprehensive range of operational conditions, providing insights into the scalability and reliability of the InTec framework. These scenarios are mentioned below.
	
	\textbf{\textit{Experiment 1:}} Explores the optimal data reduction rate to enhance network efficiency and minimize latency.
	
	\textbf{\textit{Experiment 2:}} Investigates the impact of varying data window sizes on network traffic and response times.
	
	\textbf{\textit{Experiment 3:}} Examines the effects of the number of active sensors on network traffic and latency.
	
	\textbf{\textit{Experiment 4:}} Assesses how the frequency of user requests influences network performance and service delivery.
	
	\textbf{\textit{Experiment 5:}} Assesses the performance of various frameworks, including InTec, under heavy network traffic and user demand.
	
	\subsection{\textbf{Experiment 1:} Evaluating the Effects of Data Reduction Rates on Network Efficiency and Latency Across Frameworks}
	This experiment focuses on the influence of varying data reduction rates, employing PCA and AE algorithms, on key network performance indicators. Specifically, it aims to uncover how data feature reduction by 24\% and 66\% impacts metrics such as latency, network traffic, throughput, and power consumption across
	the different layers of InTec architecture.
	
	Data reduction rates are expected to impact dependent variables because they
	directly influence the volume and complexity of data that needs to be transmitted,
	stored, and processed. Reduction in data dimensionality can lead to less network
	congestion (reduced network traffic), improved speed in data processing (lower
	latency), and potentially increased throughput as less data needs to be handled at
	any given time. Furthermore, fewer data features to analyze can result in lower
	power consumption since fewer computational resources are required. The experiment aims to quantify these effects and establish the most beneficial reduction
	rates for the InTec framework’s efficiency and effectiveness.
	
	With fixed variables like the data window size, sensor count, and user interactions, the experiment sought to observe the effects of two distinct data reduction
	rates. The setup mirrored the standardized environment, with 30 sensors transmitting data at predetermined intervals. The edge and cloud servers were emulated on virtual platforms, with all devices interconnected via a TCP/IP network to facilitate seamless data exchanges.

	\begin{landscape}
		\begin{longtable}{p{2.5cm}p{5cm}p{3.5cm}p{2.5cm}p{2cm}p{2cm}p{1.5cm}p{1.5cm}}
			\caption{Experimental design parameters} \\
			\toprule
			Exp & Question & Parameters & Reduction model & Reduction rate & Window size & Sensors & Users \\
			\midrule
			\endfirsthead
			\caption[]{Experimental design parameters (continued)} \\
			\toprule
			Exp & Question & Parameters & Reduction model & Reduction rate & Window size & Sensors & Users \\
			\midrule
			\endhead
			\bottomrule
			\endfoot
			
			Experience 1 & What reduction rate improves network traffic and response latency more effectively? & Network Traffic, Latency, Throughput, Power Consumption & PCA & Variable: 24\%, 66\% & Constant: 25 & Constant: 30 & Constant: 30 \\
			& & & AE & Variable: 24\%, 66\% & Constant: 25 & Constant: 30 & Constant: 30 \\
			\addlinespace
			Experience 2 & What window size should be used for data to reduce network traffic and latency? & Network Traffic, Latency, Throughput, Power Consumption & PCA & Constant: 66\% & Variable: 25, 50, 100 & Constant: 30 & Constant: 30 \\
			& & & AE & Constant: 66\% & Variable: 25, 50, 100 & Constant: 30 & Constant: 30 \\
			\addlinespace
			Experience 3 & How does the number of active sensors in the network affect network traffic and response delay? & Network Traffic, Latency, Throughput, Power Consumption & PCA & Constant: 66\% & Constant: 25 & Variable: 10, 20, 30 & Constant: 30 \\
			& & & AE & Constant: 66\% & Constant: 25 & Variable: 10, 20, 30 & Constant: 30 \\
			\addlinespace
			Experience 4 & How does the number of user requests in a unit of time affect network traffic and response latency? & Network Traffic, Latency, Throughput, Power Consumption & PCA & Constant: 66\% & Constant: 25 & Constant: 30 & Variable: 10, 20, 30 \\
			& & & AE & Constant: 66\% & Constant: 25 & Constant: 30 & Variable: 10, 20, 30 \\
			\addlinespace
			Experience 5 & What is the performance of evaluated frameworks under network traffic load and response latency in high-pressure conditions? & Network Traffic, Latency, Throughput, Power Consumption & PCA & Constant: 66\% & Constant: 25 & Variable: 50, 60, 70, 80, 90, 100 & Variable: 50, 60, 70, 80, 90, 100 \\
			& & & AE & Constant: 66\% & Constant: 25 & Variable: 50, 60, 70, 80, 90, 100 & Variable: 50, 60, 70, 80, 90, 100 \\
			
		\end{longtable}
	\end{landscape}
	
	Table 10 and Figs. 15 and 16 compare the AE and PCA data reduction algorithms across different metrics and reduction rates within various configurations:
	Cloud, Edge-Cloud (baseline), Edge, and the proposed InTec framework. The
	improvement percentages highlight the effectiveness of the InTec framework in
	optimizing various performance variables over traditional models. In addition,
	the results can be discussed in terms of several parameters as follows:
	
	\textit{Latency improvements:} The InTec framework’s impressive latency improvements (over 92\% across most cases) highlight its ability to optimize real-time
	processing significantly. This decrease in latency is likely due to InTec’s distributed processing approach, which minimizes the amount of data sent across the
	network by performing data reduction early in the pipeline.
	
	\textit{Network efficiency:} With reductions of approximately 11.66\% in network traffic
	and 8.56\% in throughput, the results indicate that InTec manages network congestion
	effectively. Notably, the larger data reduction rate (66\%) achieved using the PCA
	algorithm led to a greater decrease in network traffic, which aligns with the framework’s goal of alleviating data loads.
	
	\textit{Power consumption:} Power usage results across different layers reveal interesting trends. While edge and cloud layers show reduced power consumption
	(21.91\% and 28.55\% improvements, respectively), the sensor layer saw an increase of approximately 6.18\%.
	
	\textit{Data reduction:} The comparison between PCA and AE shows that PCA, particularly at a 66\% reduction rate, outperforms AE regarding latency, traffic, and throughput improvements. PCA’s superior performance likely stems from its ability to retain
	essential data features while significantly reducing data dimensions, leading to less
	network burden without compromising model interpretability. This indicates that
	PCA’s feature selection capabilities are particularly well-suited for IoT applications,
	where maintaining accuracy with reduced data volumes is crucial for real-time processing.
	
	The outcomes of Experiment 1 illustrate that the strategic integration of PCAbased data reduction within the InTec framework enables enhanced network performance and reduced latency, power consumption, and network traffic. This experiment suggests that applying such reduction techniques within a distributed IoT
	system is instrumental in overcoming the challenges posed by data-heavy applications, particularly in dynamic environments. The InTec framework, through its
	unique architecture, shows promise as a scalable solution for optimizing data flow,
	energy use, and real-time responsiveness, making it an ideal choice for IoT applications with high data processing demands.
	
	\begin{figure}[h]
		\centering
		\includegraphics[width=0.9\textwidth]{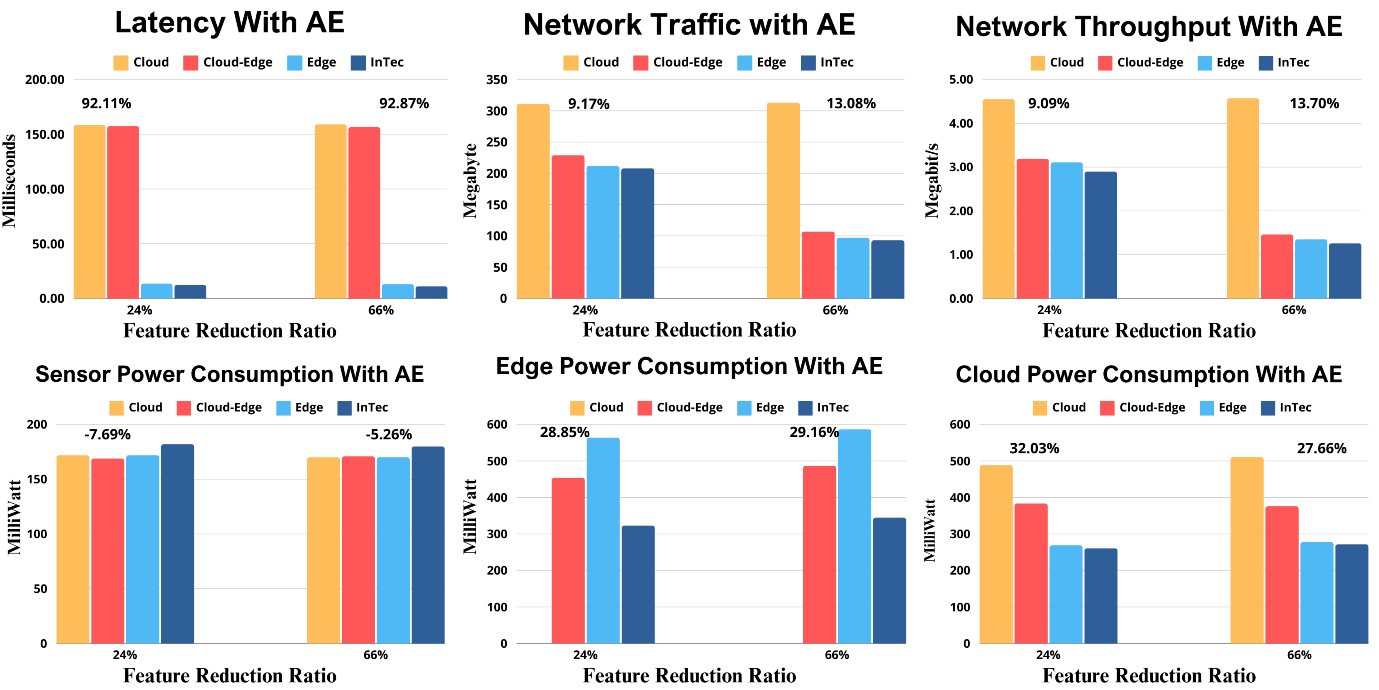}
		\caption{Charts of experiment 1 with AE algorithm}
	\end{figure}
	
	\begin{figure}[h]
		\centering
		\includegraphics[width=0.9\textwidth]{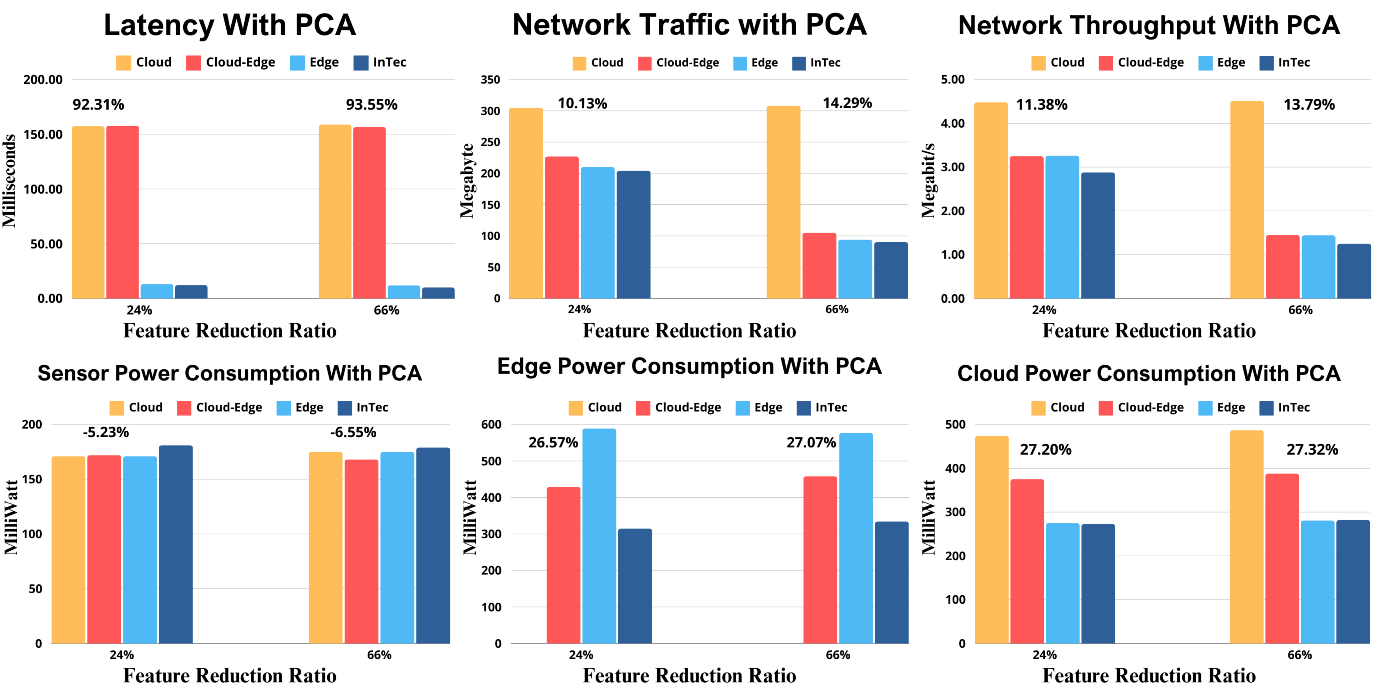}
		\caption{Charts of experiment 1 with PCA algorithm}
	\end{figure}

	\subsection{\textbf{Experiment 2:} Assessing the Influence of Sensor Data Window Size on Network Traffic and Response Times in Various Frameworks}
	Experiment 2 aimed to explore how varying data transmission window sizes (25,
	50, and 100) align with methodologies from prior research \cite{c15} and impact network
	traffic, response times, throughput, and power consumption across different frameworks. This study provides insights into optimizing data transmission strategies for
	enhanced network performance and efficiency.
	
	Adjusting the data transmission window size influences vital performance indicators due to its impact on how data is managed over the network. A larger window
	size could lead to higher network traffic volumes and potentially longer response
	times as more data is transmitted less frequently. Conversely, smaller window sizes
	may increase the frequency of transmissions, which can reduce individual data
	packet sizes, potentially enhancing response times and adding communication overhead. These variations directly affect throughput efficiency and power consumption
	at all network layers, necessitating a balance to optimize overall system performance. Experiment 2 investigates the optimal window size that harmonizes these
	factors for improved network efficiency.
	
	The experiment was structured to maintain constant variables such as data reduction rate (66\%), number of sensors (30), and number of users (30) to isolate the
	effects of changing window sizes. The experiment used AE and PCA algorithms to
	reduce data dimensions and assess their effectiveness across different window sizes.
	In addition, the results can be discussed in terms of several parameters as follows:
	
	\textit{Latency improvements:} As window sizes increased, latency improvements
	slightly decreased, with smaller window sizes (particularly a window size of 25)
	providing the most significant reduction in response times—up to 93.56\% with
	PCA. This trend likely results from the smaller windows allowing more frequent
	data updates, which aligns well with InTec’s distributed framework to handle
	real-time processing demands.
	
		\textit{Network efficiency:} The improvements in network traffic and throughput were
	most pronounced with smaller window sizes, achieving an average of 14.98\% and
	13.53\%, respectively, with PCA. Smaller windows reduce the data payload per
	transmission, leading to smoother data flow and minimizing bottlenecks within
	the network.

	\begin{landscape}
		\begin{table}[t]
			\centering
			\caption{Results of experiment 1—comparison of AE and PCA for different reduction ratios}
			\adjustbox{width=1.65\textwidth}{
				\begin{tabular}{llllllllllll}
					\toprule
					\multicolumn{2}{l}{} & \multicolumn{5}{c}{AE} & \multicolumn{5}{c}{PCA} \\
					\cmidrule(r){3-7} \cmidrule(r){8-12}
					Reduction Ratio & Variables & Cloud & \begin{tabular}[c]{@{}l@{}}Edge-Cloud \\ (base)\end{tabular} & Edge & InTec & \begin{tabular}[c]{@{}l@{}}Improve\\ \%\end{tabular} & Cloud & \begin{tabular}[c]{@{}l@{}}Edge-Cloud \\ (base)\end{tabular} & Edge & InTec & \begin{tabular}[c]{@{}l@{}}Improve\\ \%\end{tabular} \\
					\midrule
					\multirow{6}{*}{24\%} & Latency (ms) & 158.52 & 157.46 & 13.59 & 12.43 & 92.11 & 157.28 & 157.78 & 13.24 & 12.14 & 92.31 \\
					& Traffic (MB) & 311 & 229 & 212 & 208 & 9.17 & 305 & 227 & 210 & 204 & 10.13 \\
					& Throughput (Mbps) & 4.55 & 3.19 & 3.11 & 2.9 & 9.09 & 4.48 & 3.25 & 3.26 & 2.88 & 11.38 \\
					& Sensor Power (mW) & 172 & 169 & 172 & 182 & -7.69 & 171 & 172 & 171 & 181 & -5.23 \\
					& Edge Power (mW) & 0 & 454 & 564 & 323 & 28.85 & 0 & 429 & 589 & 315 & 26.57 \\
					& Cloud Power (mW) & 489 & 384 & 269 & 261 & 32.03 & 474 & 375 & 275 & 273 & 27.20 \\
					\midrule
					\multirow{6}{*}{66\%} & Latency (ms) & 159.1 & 156.57 & 13.17 & 11.17 & 92.87 & 158.81 & 156.64 & 12.01 & 10.11 & 93.55 \\
					& Traffic (MB) & 313 & 107 & 97 & 93 & 13.08 & 308 & 105 & 94 & 90 & 14.29 \\
					& Throughput (Mbps)  & 4.57 & 1.46 & 1.35 & 1.26 & 13.70 & 4.51 & 1.45 & 1.44 & 1.25 & 13.79 \\
					& Sensor Power (mW) & 170 & 171 & 170 & 180 & -5.26 & 175 & 168 & 175 & 179 & -6.55 \\
					& Edge Power (mW) & 0 & 487 & 587 & 345 & 29.16 & 0 & 458 & 577 & 334 & 27.07 \\
					& Cloud Power (mW) & 511 & 376 & 278 & 272 & 27.66 & 487 & 388 & 281 & 282 & 27.32 \\
					\bottomrule
				\end{tabular}
			}
		\end{table}
	\end{landscape}

	\textit{Power consumption:} Edge and cloud layers saw considerable reductions in
	power consumption—an average improvement of 25.46\% and 32.65\%, respectively—thanks to InTec’s distributed load-balancing. However, sensor power consumption increased slightly (by 5.7\%) due to the additional local data processing
	required for frequent, smaller transmissions. This increase in sensor power usage
	highlights an area for potential optimization, perhaps through more energy-efficient hardware or processing algorithms.
	
	\textit{Data reduction:} Across all window sizes, PCA consistently outperformed
	AE, particularly with smaller windows, reinforcing PCA’s suitability for realtime data processing in IoT frameworks. PCA’s ability to retain essential features
	with minimal data load improves both latency and network efficiency, making it
	an optimal choice for data transmission that requires high efficiency and quick
	response times.
	
	Experiment 2 demonstrates that adjusting the data transmission window size
	significantly impacts network performance, with smaller windows providing the best balance between latency, traffic, and throughput in the InTec framework. The
	PCA algorithm emerges as the most effective for data reduction, aligning well with
	InTec’s goal to manage real-time data efficiently and responsively. This experiment
	underscores the importance of fine-tuning data handling strategies—such as window
	size and data reduction techniques—to achieve optimal performance in IoT systems.
	These findings establish that InTec’s architecture can adapt to various network conditions, making it a versatile solution for the growing demands of data-driven IoT
	environments. Table 11 and Figs. 17 and 18 present the results of this experiment.
	
	\begin{figure}[h]
		\centering
		\includegraphics[width=0.9\textwidth]{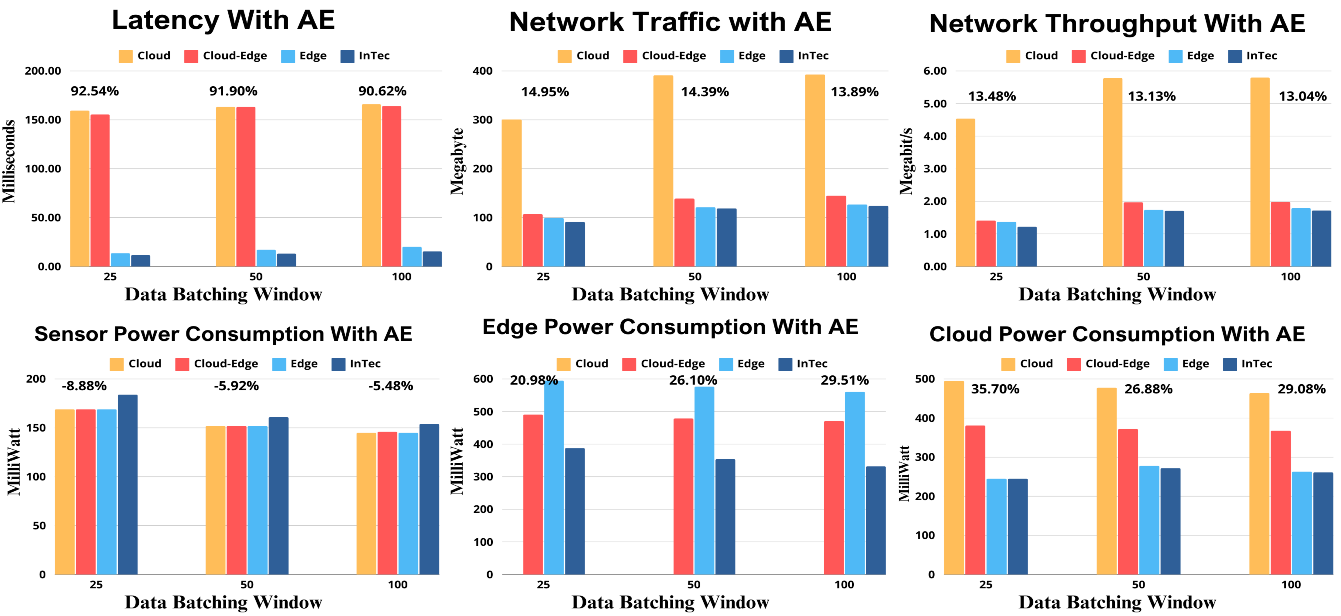}
		\caption{Charts of experiment 2 with AE algorithm}
	\end{figure}
	
	\begin{figure}[h]
		\centering
		\includegraphics[width=0.9\textwidth]{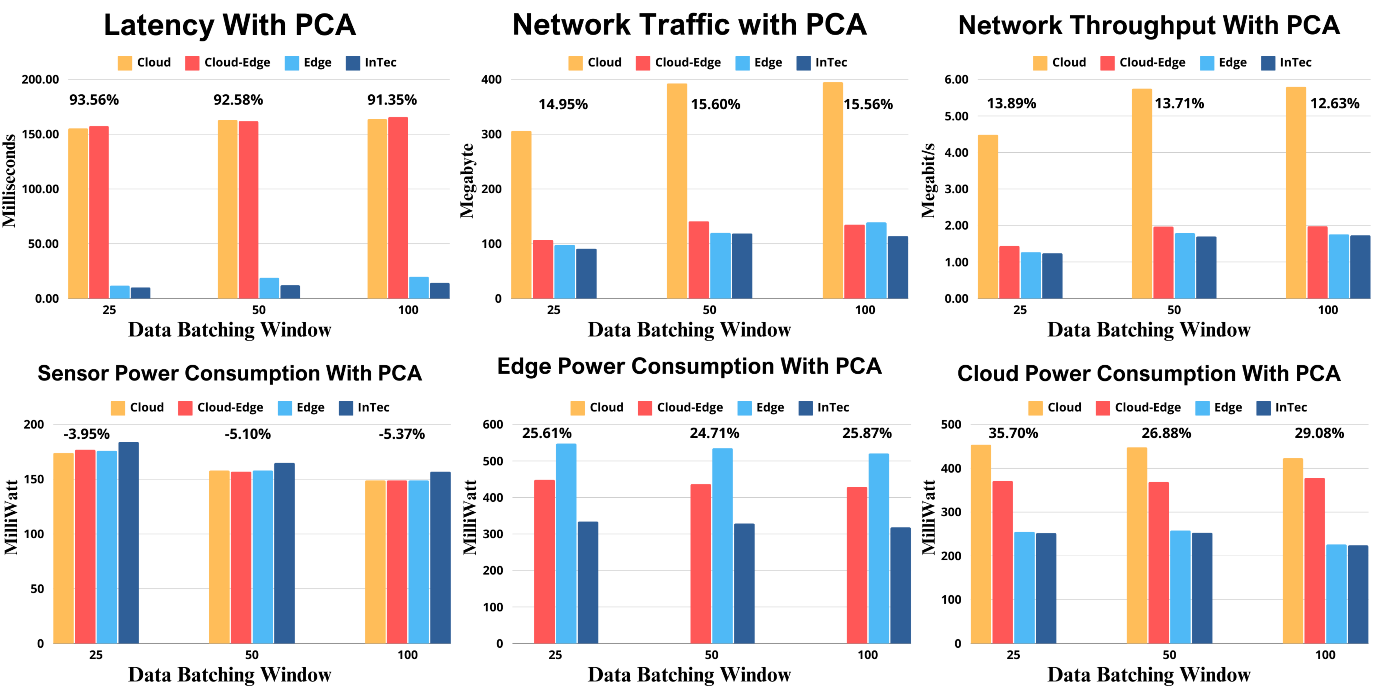}
		\caption{Charts of experiment 2 with PCA algorithm}
	\end{figure}
	
	\subsection{\textbf{Experiment 3:} Examining the Role of Active Sensor Quantity on Network Performance Metrics Across Frameworks}
	The third experiment aims to evaluate the performance of different frameworks in
	processing sensor data with varying sensor counts. It specifically investigates how
	the number of active sensors—10, 20, 30, and 40—affects critical performance indicators such as response time, network traffic, throughput, and power consumption
	across various framework layers.
	
	The count of active sensors within a network is a critical factor affecting performance metrics due to the direct correlation between the number of sensors and the
	volume of data generated. As the sensor count increases, there’s an expected uptick
	in network traffic, potentially leading to longer response times due to the higher volume of data needing processing. Conversely, a more significant number of sensors
	can lead to improved data accuracy and a richer dataset for analysis, which could
	enhance the overall throughput. This experiment assesses the impact of varying sensor counts, understanding how they influence network traffic load, response times,
	data throughput, and the consequent power consumption at the edge and cloud layers. The objective is to determine the sensor count that balances system performance
	and resource consumption.
	
	For this evaluation, the data reduction rate (66\%), window size (25), and number
	of users (10) were held constant. The experiment incrementally increased the sensor count from 10 to 40, measuring the impact on dependent variables at each step.
	The setup mirrored the initial experiment, progressively adding sensors and assessing the outcomes to determine how increasing sensor counts influences framework
	performance.
	
		The evaluative metrics are presented in Table 12 and visually illustrated in
	Figs. 19 and 20, covering aspects such as latency, network traffic, and power usage.
	This experiment provides a comprehensive view of the frameworks’ performances
	under different sensor loads. This analysis aims to unveil the scalability and robustness of each framework, with a keen emphasis on understanding the operational
	dynamics and efficiency of the InTec framework as the sensor ecosystem expands.
	In addition, the results can be discussed in terms of several parameters as follows:
	
	\textit{Latency improvements:} The InTec framework demonstrated consistent latency
	improvements across all sensor counts, maintaining above 92\% latency reduction in most cases with both AE and PCA. This result highlights InTec’s capability to handle an increasing number of sensors without compromising response time, which is essential for real-time applications.
	
	\textit{Network efficiency:} As expected, increasing sensor counts increased traffic and
	throughput demands. However, InTec managed to maintain efficiency, achieving
	traffic and throughput improvements of over 12\% and 10\%, respectively, across different sensor counts. The improvements were more prominent when using PCA,
	indicating that InTec can effectively reduce data load even as the network denser.
	
	\textit{Power consumption:} While sensor power usage slightly increased (averaging
	around a 5–10\% rise) due to the additional data processing demands, the edge and
	cloud layers showed significant energy savings, with cloud power consumption
	improving by over 27\% on average. This result suggests that the InTec framework
	is efficient in managing energy resources, effectively distributing tasks across layers to limit power drain on the cloud and edge while handling increased data input.

	\begin{landscape}
		\begin{table}[t]
			\centering
			\caption{Rsults of experiment 2—comparison of AE and PCA for different data batching windows}
			\adjustbox{width=1.65\textwidth}{
				\begin{tabular}{llllllllllll}
					\toprule
					& & \multicolumn{5}{c}{AE} & \multicolumn{5}{c}{PCA} \\
					\cmidrule(r){3-7} \cmidrule(r){8-12}
					Window Size & Variables & Cloud & \begin{tabular}[c]{@{}l@{}}Edge-Cloud \\ (base)\end{tabular} & Edge & InTec & \begin{tabular}[c]{@{}l@{}}Improve\\ \%\end{tabular} & Cloud & \begin{tabular}[c]{@{}l@{}}Edge-Cloud \\ (base)\end{tabular} & Edge & InTec & \begin{tabular}[c]{@{}l@{}}Improve\\ \%\end{tabular} \\
					\midrule
					\multirow{6}{*}{25} & Latency (ms) & 159.61 & 155.53 & 13.61 & 11.61 & 92.54 & 155.45 & 157.55 & 11.85 & 10.15 & 93.56 \\
					& Traffic (MB) & 301 & 107 & 99 & 91 & 14.95 & 306 & 107 & 98 & 91 & 14.95 \\
					& Throughput (Mbps)  & 4.54 & 1.41 & 1.37 & 1.22 & 13.48 & 4.49 & 1.44 & 1.27 & 1.24 & 13.89 \\
					& Sensor Power (mW) & 169 & 169 & 169 & 184 & -8.88 & 174 & 177 & 176 & 184 & -3.95 \\
					& Edge Power (mW) & 0 & 491 & 595 & 388 & 20.98 & 0 & 449 & 548 & 334 & 25.61 \\
					& Cloud Power (mW) & 495 & 381 & 245 & 245 & 35.70 & 454 & 371 & 255 & 252 & 32.08 \\
					\midrule
					\multirow{6}{*}{50} & Latency (ms) & 163.27 & 163.24 & 17.23 & 13.23 & 91.90 & 163.02 & 162.08 & 18.92 & 12.02 & 92.58 \\
					& Traffic (MB) & 391 & 139 & 121 & 119 & 14.39 & 393 & 141 & 120 & 119 & 15.60 \\
					& Throughput (Mbps)  & 5.78 & 1.97 & 1.74 & 1.71 & 13.20 & 5.75 & 1.97 & 1.79 & 1.7 & 13.71 \\
					& Sensor Power (mW) & 152 & 152 & 152 & 161 & -5.92 & 158 & 157 & 158 & 165 & -5.10 \\
					& Edge Power (mW) & 0 & 479 & 577 & 354 & 26.10 & 0 & 437 & 535 & 329 & 24.71 \\
					& Cloud Power (mW) & 478 & 372 & 278 & 272 & 26.88 & 448 & 369 & 258 & 253 & 31.44 \\
					\midrule
					\multirow{6}{*}{100} & Latency (ms) & 165.99 & 164.22 & 20.14 & 15.4 & 90.62 & 163.84 & 165.82 & 19.84 & 14.34 & 91.35 \\
					& Traffic (MB) & 393 & 145 & 127 & 124 & 14.48 & 395 & 135 & 139 & 114 & 15.56 \\
					& Throughput (Mbps) & 5.8 & 1.98 & 1.79 & 1.72 & 13.13 & 5.8 & 1.98 & 1.76 & 1.73 & 12.63 \\
					& Sensor Power (mW) & 145 & 146 & 145 & 154 & -5.48 & 149 & 149 & 149 & 157 & -5.37 \\
					& Edge Power (mW) & 0 & 471 & 561 & 332 & 29.51 & 0 & 429 & 521 & 318 & 25.87 \\
					& Cloud Power (mW) & 464 & 368 & 263 & 261 & 29.08 & 423 & 378 & 226 & 224 & 40.74 \\
					\bottomrule
				\end{tabular}
			}
		\end{table}
	\end{landscape}

	\textit{Data reduction:} Throughout the analysis, PCA consistently performed marginally better than AE, particularly in network efficiency and latency reduction. PCA’s
	effectiveness in reducing data dimensions while retaining key features allowed it
	to perform well even as the number of sensors increased. This result underscores
	PCA’s suitability for high-density sensor networks, where efficient data handling
	and transmission are crucial.
	
	Experiment 3 highlights the InTec framework’s scalability and robustness in handling various sensor quantities while maintaining network efficiency, low latency,
	and control power consumption. The consistent latency improvements, along with
	enhanced network traffic and throughput management, demonstrate InTec’s resilience and ability to support large IoT networks. This experiment also reinforces
	PCA’s role as the preferred data reduction algorithm in dense sensor environments,
	where network performance is critical.
	
	\begin{figure}[h]
		\centering
		\includegraphics[width=0.9\textwidth]{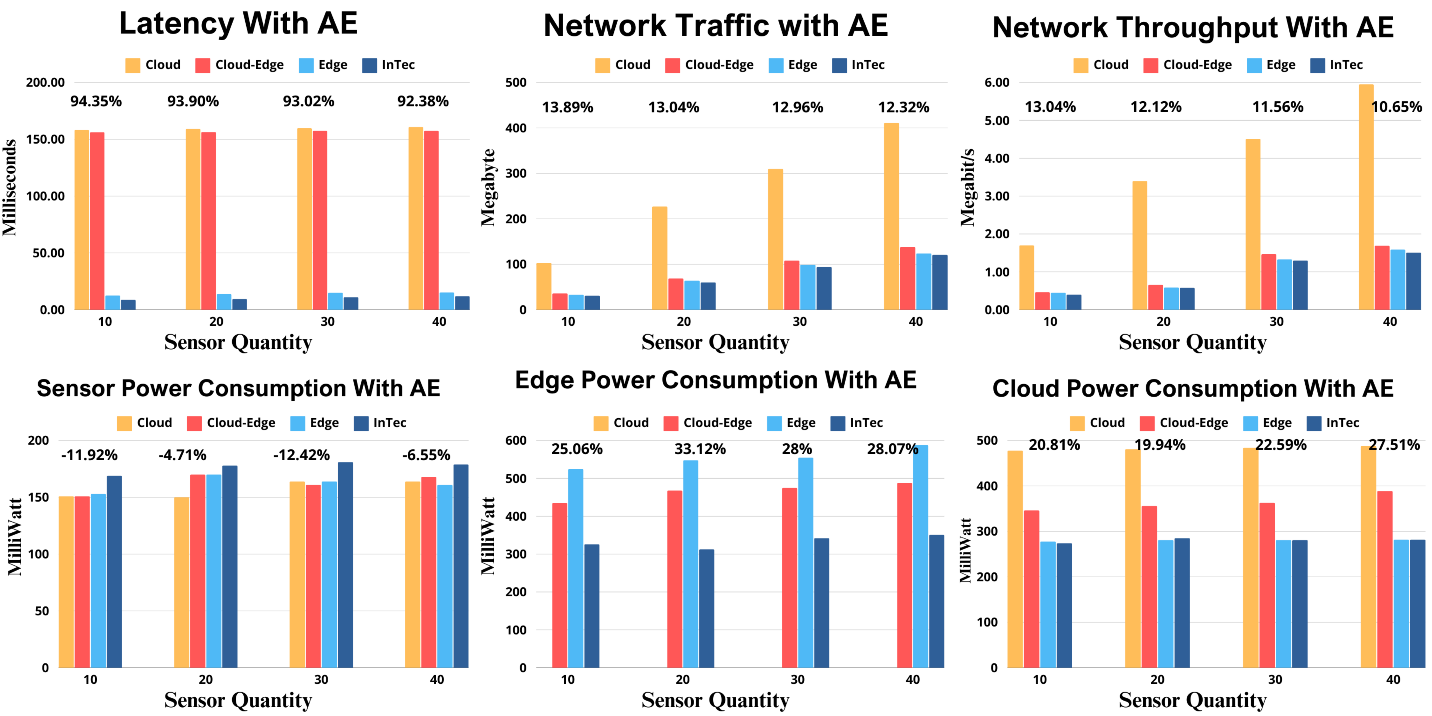}
		\caption{Charts of experiment 3 with AE algorithm}
	\end{figure}
	
	\begin{figure}[hb]
		\centering
		\includegraphics[width=0.9\textwidth]{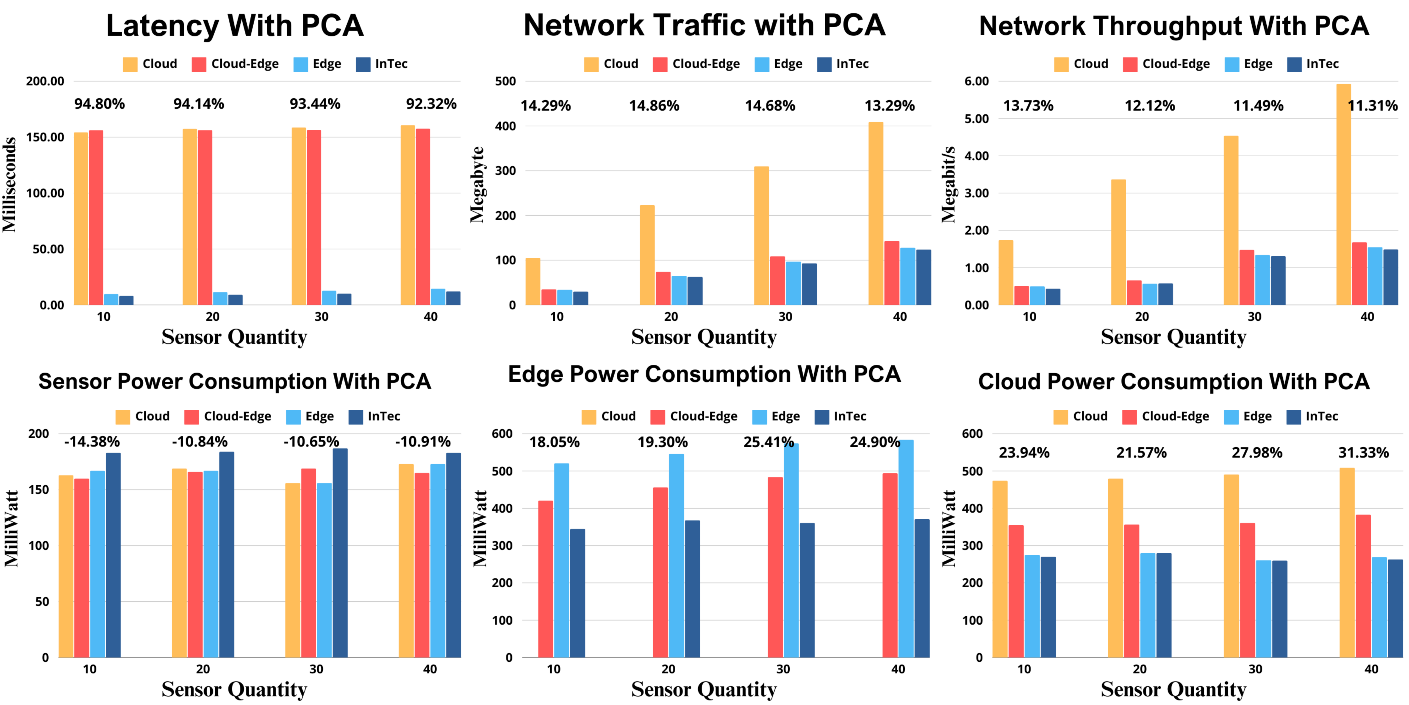}
		\caption{Charts of experiment 3 with PCA algorithm}
	\end{figure}

	\begin{landscape}
		
		\begin{table}[t]
			\centering
			\caption{Results of experiment 3—comparison of AE and PCA with varying sensor quantities}
			\adjustbox{width=1.65\textwidth}{
				\begin{tabular}{llllllllllll}
					\toprule
					& & \multicolumn{5}{c}{AE} & \multicolumn{5}{c}{PCA} \\
					\cmidrule(r){3-7} \cmidrule(r){8-12}
					Sensor Quantity & Variables & Cloud & \begin{tabular}[c]{@{}l@{}}Edge-Cloud \\ (base)\end{tabular} & Edge & InTec & \begin{tabular}[c]{@{}l@{}}Improve\\ \%\end{tabular} & Cloud & \begin{tabular}[c]{@{}l@{}}Edge-Cloud \\ (base)\end{tabular} & Edge & InTec & \begin{tabular}[c]{@{}l@{}}Improve\\ \%\end{tabular} \\
					\midrule
					\multirow{6}{*}{10} & Latency (ms) & 158.26 & 156.21 & 12.56 & 8.83 & 94.35 & 154.31 & 156.19 & 9.84 & 8.12 & 94.80 \\
					& Traffic (MB) & 103 & 36 & 33 & 31 & 13.89 & 105 & 35 & 34 & 30 & 14.29 \\
					& Throughput(Mbps) & 1.7 & 0.46 & 0.45 & 0.41 & 10.87 & 1.74 & 0.51 & 0.5 & 0.44 & 13.73 \\
					& Sensor Power (mW) & 151 & 151 & 153 & 169 & -11.92 & 163 & 160 & 167 & 183 & -14.38 \\
					& Edge Power (mW) & 0 & 435 & 525 & 326 & 25.06 & 0 & 421 & 521 & 345 & 18.05 \\
					& Cloud Power (mW) & 478 & 346 & 278 & 274 & 20.81 & 474 & 355 & 275 & 270 & 23.94 \\
					\midrule
					\multirow{6}{*}{20} & Latency (ms) & 159.18 & 156.44 & 13.78 & 9.54 & 93.90 & 157.54 & 156.34 & 11.56 & 9.16 & 94.14 \\
					& Traffic (MB) & 227 & 69 & 64 & 60 & 13.04 & 223 & 74 & 65 & 63 & 14.86 \\
					& Throughput (Mbps)  & 3.4 & 0.66 & 0.59 & 0.58 & 12.12 & 3.37 & 0.66 & 0.57 & 0.58 & 12.12 \\
					& Sensor Power (mW) & 150 & 170 & 170 & 178 & -4.71 & 169 & 166 & 167 & 184 & -10.84 \\
					& Edge Power (mW) & 0 & 468 & 548 & 313 & 33.12 & 0 & 456 & 546 & 368 & 19.30 \\
					& Cloud Power (mW) & 481 & 356 & 281 & 285 & 19.94 & 480 & 357 & 280 & 280 & 21.57 \\
					\midrule
					\multirow{6}{*}{30} & Latency (ms) & 159.8 & 157.51 & 14.99 & 10.99 & 93.02 & 158.78 & 156.49 & 12.76 & 10.26 & 93.44 \\
					& Traffic (MB) & 310 & 108 & 99 & 94 & 12.96 & 310 & 109 & 97 & 93 & 14.68 \\
					& Throughput (Mbps)  & 4.51 & 1.47 & 1.33 & 1.3 & 11.56 & 4.54 & 1.48 & 1.34 & 1.31 & 11.49 \\
					& Sensor Power (mW) & 164 & 161 & 164 & 181 & -12.42 & 156 & 169 & 156 & 187 & -10.65 \\
					& Edge Power (mW) & 0 & 475 & 555 & 342 & 28.00 & 0 & 484 & 574 & 361 & 25.41 \\
					& Cloud Power (mW) & 484 & 363 & 281 & 281 & 22.59 & 491 & 361 & 261 & 260 & 27.98 \\
					\midrule
					\multirow{6}{*}{40} & Latency (ms) & 160.89 & 157.56 & 15.21 & 12.01 & 92.38 & 160.68 & 157.64 & 14.61 & 12.11 & 92.32 \\
					& Traffic (MB) & 411 & 138 & 124 & 121 & 12.32 & 409 & 143 & 128 & 124 & 13.29 \\
					& Throughput (Mbps)  & 5.95 & 1.69 & 1.59 & 1.51 & 10.65 & 5.93 & 1.68 & 1.55 & 1.49 & 11.31 \\
					& Sensor Power (mW) & 164 & 168 & 161 & 179 & -6.55 & 173 & 165 & 173 & 183 & -10.91 \\
					& Edge Power (mW) & 0 & 488 & 588 & 351 & 28.07 & 0 & 494 & 584 & 371 & 24.90 \\
					& Cloud Power (mW) & 488 & 389 & 282 & 282 & 27.51 & 509 & 383 & 269 & 263 & 31.33 \\
					\bottomrule
				\end{tabular}
			}
		\end{table}
	\end{landscape}
	
	\subsection{\textbf{Experiment 4:} Analyzing User Request Volume's Impact on Network Traffic and Latency in Different Frameworks}
	The primary aim of Experiment 4 is to scrutinize how the volume of user requests
	impacts the operational efficiency and resource consumption of the four frameworks
	under study. This analysis focuses on understanding the frameworks’ scalability and
	responsiveness to varying levels of user engagement.
	
	The volume of user requests is a significant variable that can affect network
	efficiency and the performance of ML models within a framework. Increased user
	requests can lead to higher network traffic, potentially resulting in more significant
	latency due to the added load on the system’s processing and data transmission
	capabilities. This experiment investigates the effects of escalating user requests on
	response time, network traffic, network throughput, and power consumption within
	the evaluated frameworks. The objective is to ascertain the frameworks’ capacity to
	handle surges in user engagement without compromising on performance or efficiency, providing valuable insights into the scalability and resilience of the system
	under varying user load conditions.
	
	For a controlled comparison, the experiment maintains constant values for data
	reduction rate (66\%), window size (25), and the number of sensors (10). The variable of interest, the number of user requests, is systematically altered across four
	levels: 10, 20, 30, and 40. This setup mirrors the environment established in prior
	experiments, ensuring consistency in testing conditions. The performance metrics
	collected—response time, network traffic, throughput, and power consumption—are
	compared across these varying user volumes to draw insights into each framework’s
	capacity to handle increased demand.
	
	The outcomes of Experiment 4, as summarized in Table 13 and Figs. 21 and 22,
	reveal several critical insights into how the frameworks manage varying volumes of
	user requests. In addition, the results can be discussed in terms of several parameters
	as follows:
	
	\textit{Latency improvements:} The InTec framework demonstrates significant latency
	reductions across all user request levels, achieving over 92\% improvement in most
	cases. The highest latency improvement occurs with 40 users, reaching 92.85\%
	for PCA. This consistent performance under increased user demand indicates that
	InTec’s efficient data processing and load-balancing capabilities prevent latency
	spikes that would typically arise from higher user engagement. The framework’s
	scalability is apparent, as it can handle substantial user activity while maintaining
	responsiveness, a critical feature for real-time applications in IoT environments with
	fluctuating user loads.
	
	\textit{Network efficiency:} As the user count increases, network traffic and throughput
	demands also rise. However, the InTec framework maintains efficient network traffic
	and throughput, with improvements of 15.32\% and 15.89\% for AE and PCA, respectively, at the highest user count (40 users). The results suggest that InTec’s layered
	architecture efficiently handles increased user-generated data, preventing congestion
	and optimizing data flow. Notably, PCA’s superior performance in managing traffic
	and throughput aligns with its ability to reduce data dimensions effectively, even
	when faced with high user volumes, making it ideal for systems that require consistent network performance under variable demand.
	
	\textit{Power consumption:} Sensor power consumption slightly increases due to the
	higher processing load generated by more user requests, but significant power savings are observed at the edge and cloud layers. The most substantial energy savings
	occur in the cloud, with improvements reaching up to 34.46\% for PCA when 40
	users are active. This result highlights InTec’s energy efficiency and ability to distribute processing tasks in a way that minimizes resource strain on the cloud layer.
	
	\textit{Data reduction:} PCA consistently outperforms AE across all user request volumes, particularly in latency reduction and network efficiency. The effectiveness of
	PCA in handling high user activity without significantly impacting latency or traffic
	makes it a preferable choice for InTec when operating under heavy user loads. These
	results indicate that PCA’s dimensionality reduction technique is well-suited for scenarios with numerous simultaneous user requests, ensuring the framework remains
	efficient and responsive while conserving network and processing resources.
	
	Experiment 4 underscores InTec’s scalability and resilience in handling high user
	request volumes, an essential feature for IoT systems subject to variable demand.
	The consistent improvements in latency, network efficiency, and power consumption emphasize InTec’s ability to adapt to different user loads without compromising performance. Furthermore, the comparison between AE and PCA suggests that
	PCA is the optimal data reduction method for applications that experience high user
	engagement, as it maintains the framework’s efficiency and responsiveness. These
	findings highlight InTec’s potential for enhancing user experience and operational
	efficiency in data-intensive IoT applications, making it a scalable, resilient solution
	for diverse IoT use cases.
	
	\begin{figure}[h]
		\centering
		\includegraphics[width=0.9\textwidth]{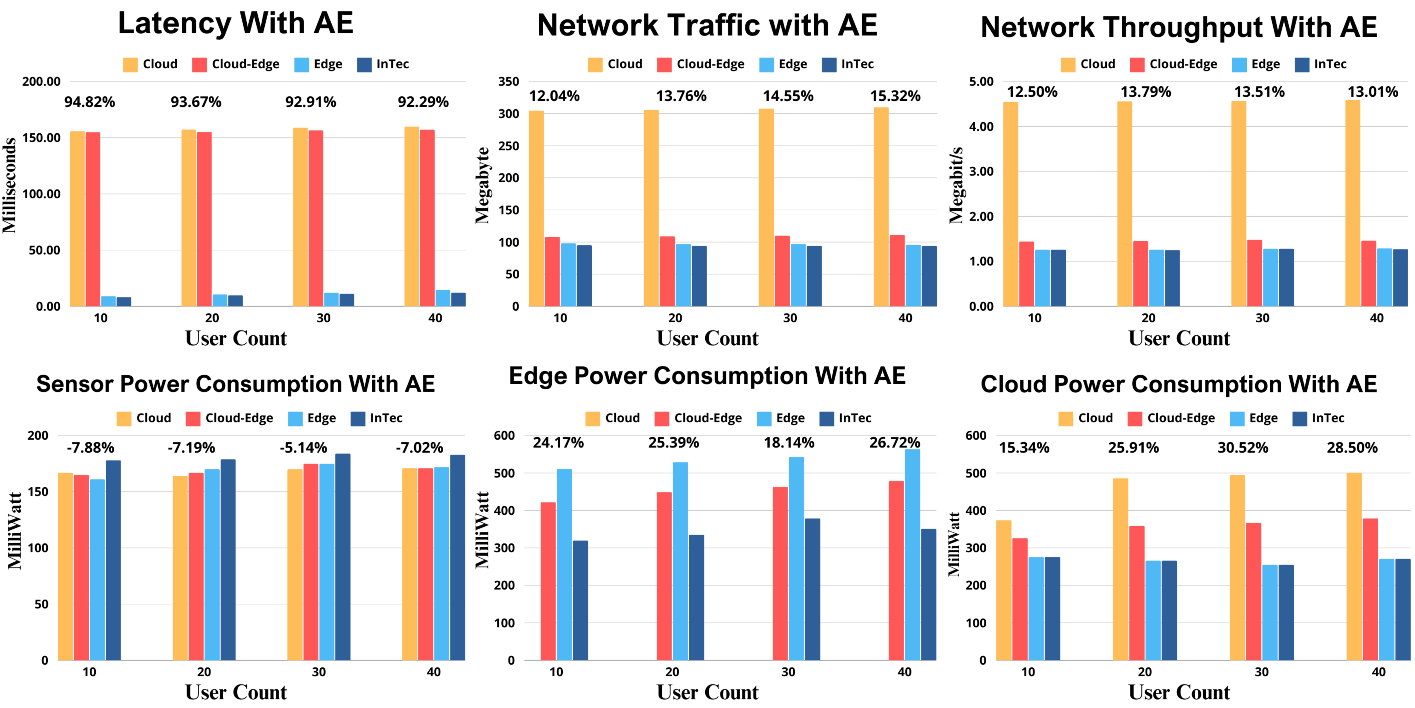}
		\caption{Charts of experiment 4 with AE algorithm}
	\end{figure}
	
	\begin{figure}[hb]
		\centering
		\includegraphics[width=0.9\textwidth]{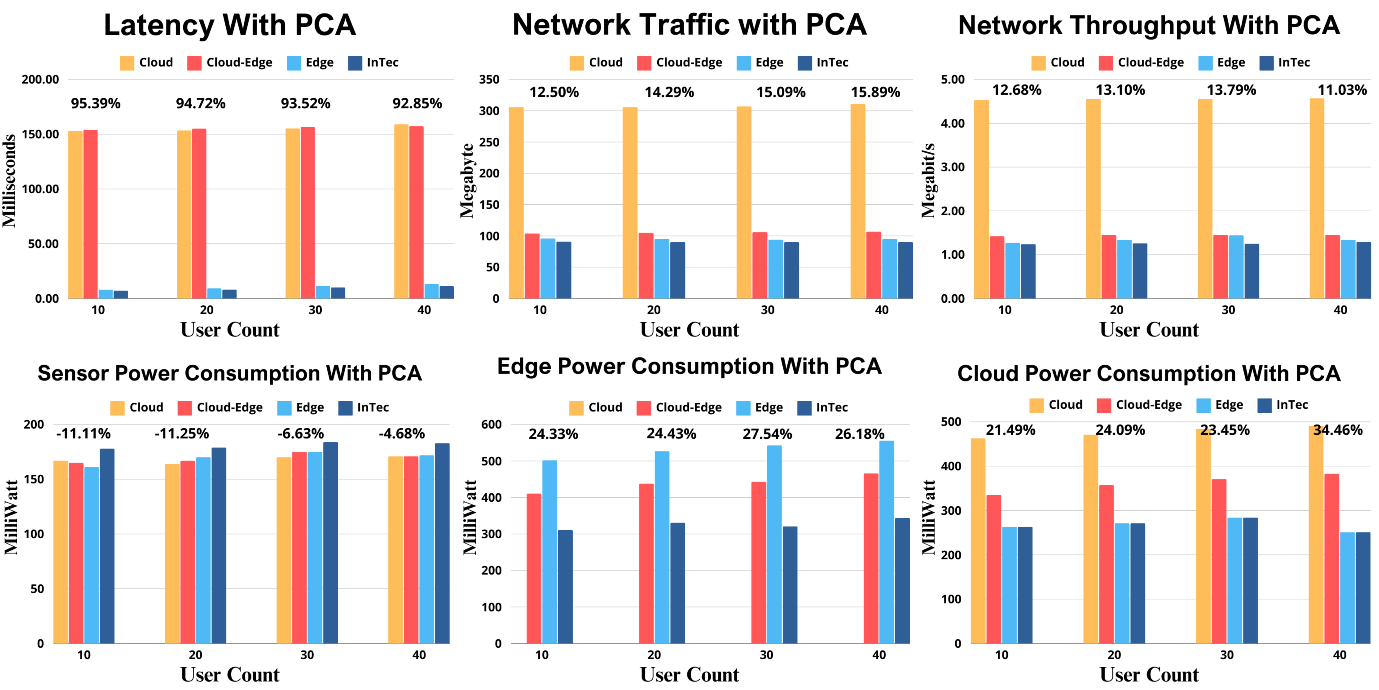}
		\caption{Charts of experiment 4 with PCA algorithm}
	\end{figure}
	
	\begin{landscape}
		
		\begin{table}[t]
			\centering
			\caption{Results of experiment 4—comparison of AE and PCA with varying user counts}
			\adjustbox{width=1.65\textwidth}{
				\begin{tabular}{llllllllllll}
					\toprule
					& & \multicolumn{5}{c}{AE} & \multicolumn{5}{c}{PCA} \\
					\cmidrule(r){3-7} \cmidrule(r){8-12}
					User Count & Variables & Cloud & \begin{tabular}[c]{@{}l@{}}Edge-Cloud \\ (base)\end{tabular} & Edge & InTec & \begin{tabular}[c]{@{}l@{}}Improve\\ \%\end{tabular} & Cloud & \begin{tabular}[c]{@{}l@{}}Edge-Cloud \\ (base)\end{tabular} & Edge & InTec & \begin{tabular}[c]{@{}l@{}}Improve\\ \%\end{tabular} \\
					\midrule
					
					\multirow{6}{*}{10} & Latency (ms) & 155.87 & 154.89 & 9.02 & 8.02 & 94.82 & 153.01 & 154.11 & 8.06 & 7.11 & 95.39 \\
					& Traffic (MB) & 305 & 108 & 98 & 95 & 12.04 & 306 & 104 & 96 & 91 & 12.50 \\
					& Throughput (Mbps)   & 4.55 & 1.44 & 1.26 & 1.26 & 12.50 & 4.53 & 1.42 & 1.27 & 1.24 & 12.68 \\
					& Sensor Power (mW) & 167 & 165 & 161 & 178 & -7.88 & 164 & 162 & 165 & 180 & -11.11 \\
					& Edge Power (mW) & 0 & 422 & 511 & 320 & 24.17 & 0 & 411 & 502 & 311 & 24.33 \\
					& Cloud Power (mW) & 374 & 326 & 276 & 276 & 15.34 & 463 & 335 & 263 & 263 & 21.49 \\
					\midrule
					\multirow{6}{*}{20} & Latency (ms) & 157.32 & 155.13 & 10.62 & 9.82 & 93.67 & 153.57 & 155.12 & 9.17 & 8.19 & 94.72 \\
					& Traffic (MB) & 306 & 109 & 97 & 94 & 13.76 & 306 & 105 & 95 & 90 & 14.29 \\
					& Throughput (Mbps)  & 4.56 & 1.45 & 1.26 & 1.25 & 13.79 & 4.56 & 1.45 & 1.34 & 1.26 & 13.10 \\
					& Sensor Power (mW) & 164 & 167 & 170 & 179 & -7.19 & 162 & 160 & 167 & 178 & -11.25 \\
					& Edge Power (mW) & 0 & 449 & 529 & 335 & 25.39 & 0 & 438 & 527 & 331 & 24.43 \\
					& Cloud Power (mW) & 486 & 359 & 266 & 266 & 25.91 & 471 & 357 & 271 & 271 & 24.09 \\
					\midrule
					\multirow{6}{*}{30} & Latency (ms) & 159.1 & 156.61 & 12.1 & 11.1 & 92.91 & 155.32 & 156.56 & 11.35 & 10.15 & 93.52 \\
					& Traffic (MB) & 308 & 110 & 97 & 94 & 14.55 & 307 & 106 & 94 & 90 & 15.09 \\
					& Throughput (Mbps)  & 4.57 & 1.48 & 1.28 & 1.28 & 13.51 & 4.56 & 1.45 & 1.44 & 1.25 & 13.79 \\
					& Sensor Power (mW) & 170 & 175 & 175 & 184 & -5.14 & 167 & 166 & 167 & 177 & -6.63 \\
					& Edge Power (mW) & 0 & 463 & 543 & 379 & 18.14 & 0 & 443 & 543 & 321 & 27.54 \\
					& Cloud Power (mW) & 495 & 367 & 255 & 255 & 30.52 & 484 & 371 & 284 & 284 & 23.45 \\
					\midrule
					\multirow{6}{*}{40} & Latency (ms) & 160.03 & 157.27 & 14.63 & 12.13 & 92.29 & 159.15 & 157.36 & 13.25 & 11.25 & 92.85 \\
					& Traffic (MB) & 310 & 111 & 96 & 94 & 15.32 & 311 & 107 & 95 & 90 & 15.89 \\
					& Throughput (Mbps)  & 4.59 & 1.46 & 1.29 & 1.27 & 13.01 & 4.57 & 1.45 & 1.34 & 1.29 & 11.03 \\
					& Sensor Power (mW) & 171 & 171 & 172 & 183 & -7.02 & 168 & 171 & 163 & 179 & -4.68 \\
					& Edge Power (mW) & 0 & 479 & 564 & 351 & 26.72 & 0 & 466 & 556 & 344 & 26.18 \\
					& Cloud Power (mW) & 501 & 379 & 271 & 271 & 28.50 & 491 & 383 & 251 & 251 & 34.46 \\
					\bottomrule
					
				\end{tabular}
			}
		\end{table}
	\end{landscape}
	
	\subsection{\textbf{Experiment 5:} Gauging Framework Performance Under High-Load Conditions: A Study on Network Traffic and Latency}
	The core aim of this experiment is to gauge the operational efficiency of various IoT
	frameworks when subjected to the highest levels of user and sensor load. The experiment assesses the performance when the number of active sensors directly matches
	the volume of user requests, scaling from 50 to 100 in increments of 10.
	
	Maintaining constant data reduction rates and window sizes, the experiment
	dynamically adjusts the counts of both sensors and users. This approach emulates a
	high-demand scenario within human activity recognition, testing each framework’s
	resilience under peak operational strain. The emulation environment replicates the
	setup used in prior experiments to ensure comparability.
	
	As depicted in Table 14 and Figs. 23 and 24, the comprehensive analysis conducted under extreme load conditions offers vital insights into the operational effectiveness of the evaluated frameworks. The examination sheds light on the frameworks’ resilience, efficiency, and adaptability when subjected to the highest degrees
	of processing demand, characterized by simultaneous heavy sensor activity and user
	request volumes. These findings provide a benchmark for understanding the scalability and robustness of IoT frameworks in real-world, high-demand scenarios, setting a foundation for future optimizations and enhancements.
	
	\textit{Latency improvements:} The InTec framework demonstrated significant latency
	improvements even under extreme load conditions, achieving an average latency
	improvement of 84.59\% with both AE and PCA at the highest load (100 sensors—100 users). This marked reduction indicates InTec’s ability to sustain efficient
	response times despite substantial data volume and processing demand.
	
	\textit{Network efficiency (traffic and throughput):} Network traffic and throughput saw
	modest improvements of 6.54\% and 1.53\%, respectively. Although these gains are
	lower than lighter load scenarios, they reveal that InTec can still manage data efficiently under heavy loads, ensuring steady data transmission and reducing network
	strain. PCA outperformed AE in reducing network traffic slightly, indicating that
	PCA’s feature extraction capabilities help alleviate data volume, which is critical
	when both sensor and user activity are high.
	
	\textit{Power consumption:} While edge and cloud layers benefited from power reductions (17.35\% at the edge and up to 39.4\% in the cloud), sensor power usage
	increased by 10.47\% due to the additional processing burden from high data loads.
	This increase in sensor power consumption emphasizes the need for optimized sensor algorithms and potential hardware improvements to manage energy more effectively under peak loads. However, the substantial power savings at the cloud and
	edge layers indicate that InTec’s distributed architecture mitigates power strain on
	these layers, supporting efficient high-load operations in energy-constrained IoT
	systems.
	
	\textit{Data reduction:} Under extreme load, PCA showed slightly better performance
	than AE, particularly in managing network traffic and throughput. PCA’s effectiveness in handling large data volumes without significantly impacting system latency
	makes it a suitable choice for high-stress scenarios. This result suggests that PCA is
	more capable of sustaining system performance and network efficiency under conditions with heavy user and sensor interactions, validating its application in dense IoT
	environments where data processing demands are high.
	
	The results from
	Experiment 5 confirm the InTec framework’s resilience and robustness under highload conditions, with substantial improvements in latency, moderate gains in network efficiency, and significant power savings at the cloud and edge levels. The findings underscore InTec’s capacity to manage large-scale IoT applications efficiently,
	particularly in scenarios where high sensor activity coincides with increased user
	demand. Furthermore, the comparison between AE and PCA suggests that PCA is
	the preferred data reduction technique for maintaining network performance and
	power efficiency under extreme operational loads, making it advantageous for highdensity IoT deployments.
	
	\begin{figure}[h]
		\centering
		\includegraphics[width=0.9\textwidth]{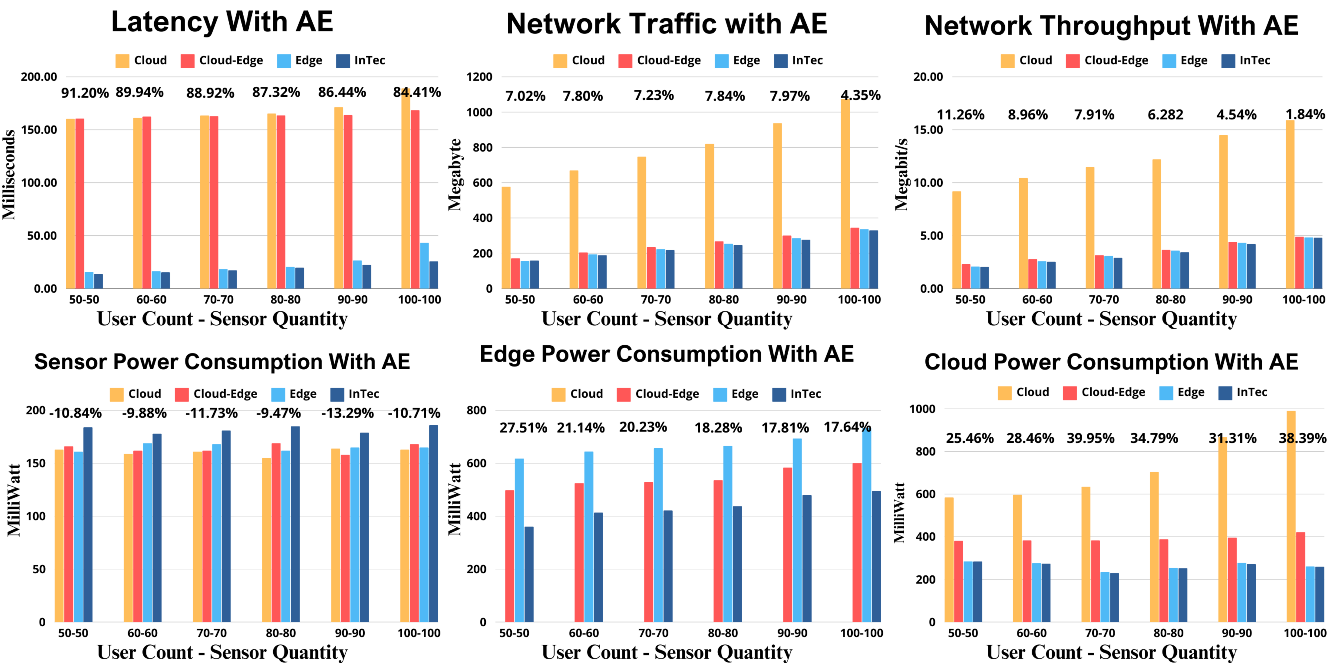}
		\caption{Charts of experiment 5 with AE algorithm}
	\end{figure}
	
	\begin{figure}[h]
		\centering
		\includegraphics[width=0.9\textwidth]{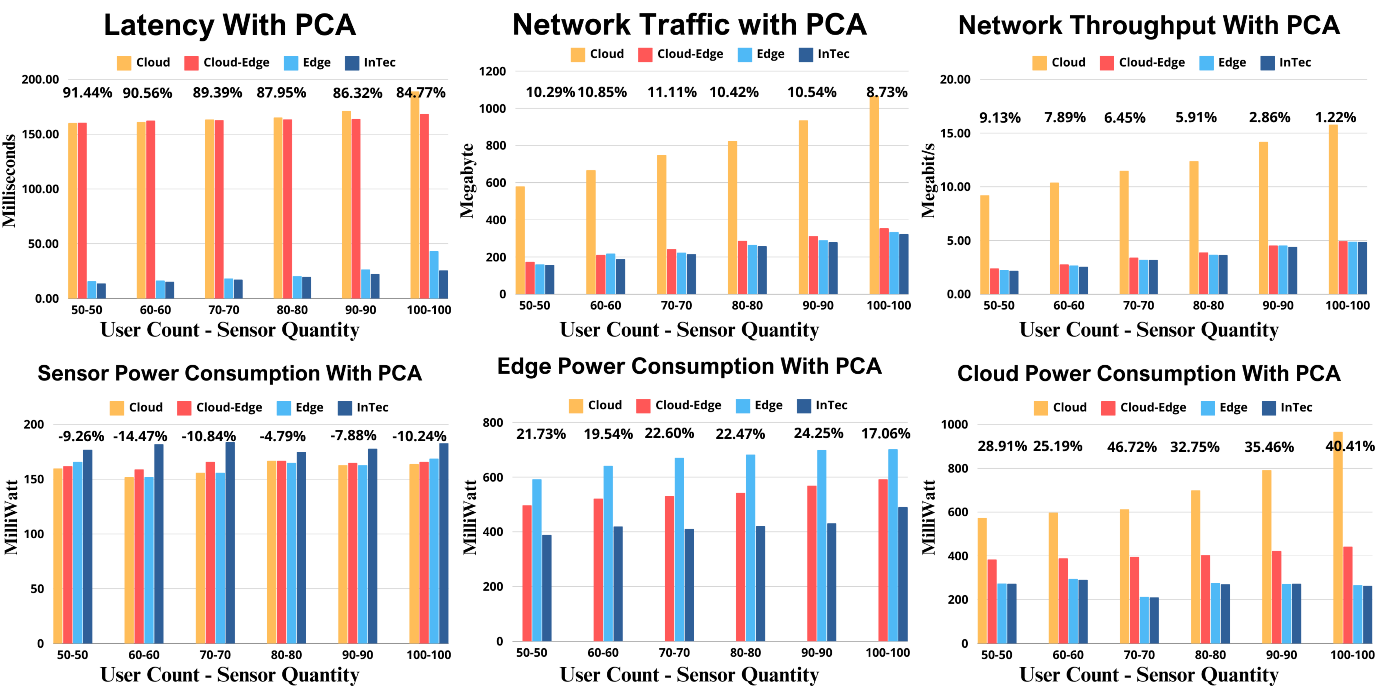}
		\caption{Charts of experiment 5 with PCA algorithm}
	\end{figure}
	
	\section{Discussions}
	The InTec framework introduces a distributed approach to handling ML pipelines
	across the Cloud, Edge, and Things layers to enhance system responsiveness and
	efficiency. This discussion delves into the experimental results, highlighting the
	framework’s strengths and areas for improvement, backed by the data from comparative experiments. Table 15 presents the average improvement independent variables in the proposed framework compared to other frameworks in each experiment
	separately. We will now delve into a more detailed examination and summary of the
	results

	\begin{landscape}
		\begin{longtable}{@{}lllllllllllll@{}}
			\caption{Results of experiment 5—comparison between AE and PCA with varying sensor quantities and user counts} \\
			\toprule
			\multicolumn{2}{c}{} & & \multicolumn{5}{c}{AE} & \multicolumn{5}{c}{PCA} \\
			\cmidrule(r){4-8} \cmidrule(r){9-13}
			\multicolumn{1}{c}{Sensor} & \multicolumn{1}{c}{User} & Variables & Cloud & \begin{tabular}[c]{@{}l@{}}Edge-Cloud \\ (base)\end{tabular} & Edge & InTec & \begin{tabular}[c]{@{}l@{}}Improve\\ \%\end{tabular} & Cloud & \begin{tabular}[c]{@{}l@{}}Edge-Cloud \\ (base)\end{tabular} & Edge & InTec & \begin{tabular}[c]{@{}l@{}}Improve\\ \%\end{tabular} \\
			\midrule
			\endfirsthead
			\caption[]{Experiment 5 Results: Results of experiment 5—comparison between AE and PCA with varying sensor quantities and user counts (continued)} \\
			\toprule
			\multicolumn{2}{c}{} & & \multicolumn{5}{c}{AE} & \multicolumn{5}{c}{PCA} \\
			\cmidrule(r){4-8} \cmidrule(r){9-13}
			\multicolumn{1}{c}{Sensor} & \multicolumn{1}{c}{User} & Variables & Cloud & \begin{tabular}[c]{@{}l@{}}Edge-Cloud \\ (base)\end{tabular} & Edge & InTec & \begin{tabular}[c]{@{}l@{}}Improve\\ \%\end{tabular} & Cloud & \begin{tabular}[c]{@{}l@{}}Edge-Cloud \\ (base)\end{tabular} & Edge & InTec & \begin{tabular}[c]{@{}l@{}}Improve\\ \%\end{tabular} \\
			\midrule
			\endhead
			\endfoot
			
			\multirow{6}{*}{50} & \multirow{6}{*}{50} & Latency (ms) & 162.61 & 161.21 & 18.11 & 14.18 & 91.20 & 160.35 & 160.56 & 15.75 & 13.75 & 91.44 \\
			& & Traffic (MB) & 578 & 171 & 156 & 159 & 7.02 & 580 & 175 & 161 & 157 & 10.29 \\
			& & Throughput (Mbps) & 9.2 & 2.31 & 2.09 & 2.05 & 11.26 & 9.22 & 2.41 & 2.25 & 2.19 & 9.13 \\
			& & Sensor Power (mW) & 163 & 166 & 161 & 184 & -10.84 & 160 & 162 & 166 & 177 & -9.26 \\
			& & Edge Power (mW) & 0 & 498 & 618 & 361 & 27.51 & 0 & 497 & 592 & 389 & 21.73 \\
			& & Cloud Power (mW) & 585 & 381 & 285 & 284 & 25.46 & 574 & 384 & 274 & 273 & 28.91 \\
			\midrule
			\multirow{6}{*}{60} & \multirow{6}{*}{60} & Latency (ms) & 164.85 & 162.26 & 19.95 & 16.32 & 89.94 & 161.24 & 162.58 & 16.44 & 15.34 & 90.56 \\
			& & Traffic (MB) & 670 & 205 & 194 & 189 & 7.80 & 668 & 212 & 219 & 189 & 10.85 \\
			& & Throughput (Mbps) & 10.45 & 2.79 & 2.59 & 2.54 & 8.96 & 10.4 & 2.79 & 2.67 & 2.57 & 7.89 \\
			& & Sensor Power (mW) & 159 & 162 & 169 & 178 & -9.88 & 152 & 159 & 152 & 182 & -14.47 \\
			& & Edge Power (mW) & 0 & 525 & 645 & 414 & 21.14 & 0 & 522 & 642 & 420 & 19.54 \\
			& & Cloud Power (mW) & 597 & 383 & 276 & 274 & 28.46 & 598 & 389 & 295 & 291 & 25.19 \\
			\midrule
			\multirow{6}{*}{70} & \multirow{6}{*}{70} & Latency (ms) & 165.96 & 162.98 & 21.06 & 18.06 & 88.92 & 163.58 & 162.94 & 18.28 & 17.28 & 89.39 \\
			& & Traffic (MB) & 748 & 235 & 223 & 218 & 7.23 & 749 & 243 & 224 & 216 & 11.11 \\
			& & Throughput (Mbps) & 11.48 & 3.16 & 3.08 & 2.91 & 7.91 & 11.5 & 3.41 & 3.21 & 3.19 & 6.45 \\
			& & Sensor Power (mW) & 161 & 162 & 168 & 181 & -11.73 & 156 & 166 & 156 & 184 & -10.84 \\
			& & Edge Power (mW) & 0 & 529 & 657 & 422 & 20.23 & 0 & 531 & 671 & 411 & 22.60 \\
			& & Cloud Power (mW) & 635 & 383 & 235 & 230 & 39.95 & 613 & 396 & 213 & 211 & 46.72 \\
			\midrule
			\multirow{6}{*}{80} & \multirow{6}{*}{80} & Latency (ms) & 168.43 & 163.57 & 23.44 & 20.74 & 87.32 & 165.43 & 163.68 & 20.43 & 19.73 & 87.95 \\
			& & Traffic (MB) & 820 & 268 & 264 & 252 & 5.97 & 824 & 288 & 291 & 286 & 0.69 \\
			& & Throughput (Mbps) & 12.2 & 3.66 & 3.59 & 3.43 & 6.28 & 12.4 & 3.89 & 3.69 & 3.66 & 5.91 \\
			& & Sensor Power (mW) & 155 & 169 & 162 & 185 & -9.47 & 167 & 167 & 165 & 175 & -4.79 \\
			& & Edge Power (mW) & 0 & 536 & 666 & 438 & 18.28 & 0 & 543 & 683 & 421 & 22.47 \\
			& & Cloud Power (mW) & 704 & 388 & 254 & 253 & 34.79 & 700 & 403 & 276 & 271 & 32.75 \\
			\midrule
			\pagebreak	
			\multirow{6}{*}{90} & \multirow{6}{*}{90} & Latency (ms) & 190.23 & 164.33 & 46.28 & 22.28 & 86.44 & 171.34 & 163.99 & 26.64 & 22.44 & 86.32 \\
			& & Traffic (MB) & 938 & 301 & 286 & 277 & 7.97 & 935 & 313 & 291 & 280 & 10.54 \\
			& & Throughput (Mbps) & 14.5 & 4.41 & 4.33 & 4.21 & 4.54 & 14.2 & 4.54 & 4.54 & 4.41 & 2.86 \\
			& & Sensor Power (mW) & 164 & 158 & 165 & 179 & -13.29 & 163 & 165 & 163 & 178 & -7.88 \\
			& & Edge Power (mW) & 0 & 584 & 694 & 480 & 17.81 & 0 & 569 & 699 & 431 & 24.25 \\
			& & Cloud Power (mW) & 868 & 396 & 278 & 272 & 31.31 & 792 & 423 & 272 & 273 & 35.46 \\
			\midrule
			\multirow{6}{*}{100} & \multirow{6}{*}{100} & Latency (ms) & 200.35 & 168.33 & 52.15 & 26.25 & 84.41 & 189.22 & 168.49 & 43.26 & 25.66 & 84.77 \\
			& & Traffic (MB) & 1075 & 345 & 336 & 330 & 4.35 & 1067 & 355 & 335 & 324 & 8.73 \\
			& & Throughput (Mbps)  & 15.91 & 4.9 & 4.84 & 4.81 & 1.84 & 15.8 & 4.93 & 4.9 & 4.87 & 1.22 \\
			& & Sensor Power (mW) & 163 & 168 & 165 & 186 & -10.71 & 164 & 166 & 169 & 183 & -10.24 \\
			& & Edge Power (mW) & 0 & 601 & 734 & 495 & 17.64 & 0 & 592 & 702 & 491 & 17.06 \\
			& & Cloud Power (mW) & 991 & 422 & 261 & 260 & 38.39 & 967 & 443 & 267 & 264 & 40.41 \\
			\bottomrule
		\end{longtable}
	\end{landscape}

		\textit{Latency Improvements:} InTec consistently reduced latency across all scenarios,
	achieving an average latency improvement of 85.89\% over cloud-based models,
	16.99\% over edge-only setups, and 81.56\% over the baseline Edge-Cloud framework. These reductions stem from InTec’s localized data processing capabilities,
	which allow for preliminary data analysis and preprocessing close to the source,
	decreasing reliance on cloud communication. This localized processing reduces typical delays associated with cloud-based frameworks and supports real-time applications in IoT environments. Such enhancements highlight the advantages of InTec’s
	multi-layered structure, particularly in settings requiring prompt responses.
	
	\textit{Network Efficiency:} Network traffic and throughput showed significant gains due
	to InTec’s data reduction techniques applied at the edge layer. By reducing data volume before transmission, the framework lessens network load, resulting in a 62.50\%
	improvement in network traffic and a 65.08\% increase in throughput compared to
	cloud-centric frameworks. The results are more pronounced in high-data scenarios,
	where efficient data handling minimizes unnecessary data transfers. This outcome suggests that strategic data reduction is integral for sustaining network efficiency in
	bandwidth-constrained IoT systems.

	\textit{Power Consumption:} The InTec framework demonstrated substantial power savings at the edge and cloud layers despite a slight increase (7.5\%) in sensor-level
	power consumption due to local data processing demands. Edge energy consumption improved by 35.47\% over edge-based frameworks, while cloud power usage saw
	a 44.54\% reduction compared to traditional cloud-only processing. These improvements underscore InTec’s effectiveness in redistributing computational tasks, significantly reducing energy requirements at higher layers by shifting processing to the
	sensor level.
	
	This trade-off between increased sensor power consumption and substantial
	energy savings at the edge and cloud layers illustrates a critical balance in the InTec framework. The increase in sensor energy usage is offset by the dramatic reductions
	in latency, network traffic, and power consumption at higher architectural layers. By
	redistributing computational tasks to the sensors, the framework reduces the burden on the edge and cloud, leading to overall system efficiency. This strategic tradeoff emphasizes the importance of optimizing energy use across the entire system,
	acknowledging that localized increases in energy consumption can result in greater
	overall energy savings and performance improvements.
	
	\textit{Data Reduction:} PCA outperformed AE regarding response time and network
	efficiency, proving advantageous in scenarios where linear dimensionality reduction
	maintains data integrity. This suggests that selecting the most suitable data reduction algorithm based on the specific IoT application and data characteristics can further optimize InTec’s performance. The findings recommend that IoT applications
	benefiting from linear reduction techniques, like PCA, can achieve superior network
	and processing efficiency. In contrast, applications with non-linear data patterns may
	require alternative reduction methods to maximize performance.
	
	\textit{Scalability of the InTec Framework:} Evaluating the InTec framework’s effectiveness includes assessing its scalability, which refers to its ability to handle increasing workloads, a more significant number of devices, and higher data volumes without compromising performance or requiring significant architectural changes. The
	framework demonstrates strong potential for both horizontal and vertical scalability. Horizontal scalability (scaling out/in) allows for adding more IoT devices, edge
	servers, and cloud resources as needed, which is crucial for IoT applications with
	exponential device growth. The distributed nature of the framework supports scaling out by adding more edge nodes to manage increased data effectively. Vertical
	scalability (scaling up/down) involves adding more power (CPU, RAM) to the existing infrastructure, allowing edge servers to handle more sophisticated data processing tasks or serve more IoT devices. This is important for data-intensive applications requiring significant computational resources for real-time analytics. Thus, the
	InTec framework exhibits a strong foundation for scalability, which is essential for
	the dynamic and growing nature of IoT applications.
	
	\begin{table}[t]
		\centering
		\caption{Comparative analysis of the InTec framework based on average experimental results}
		\adjustbox{width=\textwidth}{
			\begin{tabular}{@{}lllll@{}}
				\toprule
				Experiments & Variables & \begin{tabular}[c]{@{}l@{}}In comparison to \\ the Cloud framework\end{tabular} & \begin{tabular}[c]{@{}l@{}}In comparison to \\ the Edge framework\end{tabular} & \begin{tabular}[c]{@{}l@{}}In comparison to \\ the Edge-Cloud (base) framework\end{tabular} \\
				\midrule
				\multirow{6}{*}{Experiment 1} & Latency (ms) & 74.21 & 9.57 & 74.16 \\
				& Traffic (MB) & 41.46 & 2.62 & 9.33 \\
				& Throughput (Mbps) & 43.34 & 7.65 & 9.59 \\
				& Sensor Power (mW) & -3.97 & -3.97 & -4.95 \\
				& Edge Power (mW) & 0 & 34.52 & 22.33 \\
				& Cloud Power (mW) & 35.58 & 1.1 & 22.84 \\
				\midrule
				\multirow{6}{*}{Experiment 2} & Latency (ms) & 78.96 & 20 & 78.94 \\
				& Traffic (MB) & 59.84 & 5.44 & 12.85 \\
				& Throughput (Mbps) & 60.98 & 3.67 & 11.43 \\
				& Sensor Power (mW) & -5.22 & -5.05 & -4.96 \\
				& Edge Power (mW) & 0 & 32.97 & 21.83 \\
				& Cloud Power (mW) & 38.92 & 0.99 & 27.99 \\
				\midrule
				\multirow{6}{*}{Experiment 3} & Latency (ms) & 93.63 & 22.89 & 83.15 \\
				& Traffic (MB) & 70.82 & 5.23 & 12.15 \\
				& Throughput (Mbps) & 76.02 & 4.28 & 10.43 \\
				& Sensor Power (mW) & -12.11 & -10.27 & -9.15 \\
				& Edge Power (mW) & 0 & 37.43 & 22.43 \\
				& Cloud Power (mW) & 43.46 & 0.56 & 21.74 \\
				\midrule
				\multirow{6}{*}{Experiment 4} & Latency (ms) & 93.81 & 11.51 & 83.35 \\
				& Traffic (MB) & 69.99 & 3.92 & 12.6 \\
				& Throughput (Mbps) & 72.32 & 3.45 & 11.49 \\
				& Sensor Power (mW) & -7.89 & -7.36 & -6.77 \\
				& Edge Power (mW) & 0 & 37.03 & 21.88 \\
				& Cloud Power (mW) & 42.72 & 0 & 22.64 \\
				\midrule
				\multirow{6}{*}{Experiment 5} & Latency (ms) & 88.84 & 20.99 & 88.22 \\
				& Traffic (MB) & 70.41 & 3.41 & 7.71 \\
				& Throughput (Mbps) & 72.76 & 2.38 & 6.19 \\
				& Sensor Power (mW) & -12.81 & -10.88 & -10.28 \\
				& Edge Power (mW) & 0 & 35.43 & 20.85 \\
				& Cloud Power (mW) & 62.06 & 0.95 & 33.98 \\
				\midrule
				\multirow{6}{*}{Average Results} & Latency (ms) & 85.89 & 16.992 & 81.564 \\
				& Traffic (MB) & 62.504 & 4.124 & 10.928 \\
				& Throughput (Mbps) & 65.084 & 4.286 & 9.826 \\
				& Sensor Power (mW) & -8.4 & -7.506 & -7.222 \\
				& Edge Power (mW) & 0 & 35.476 & 21.864 \\
				& Cloud Power (mW) & 44.548 & 0.72 & 25.838 \\
				\bottomrule
			\end{tabular}
		}
	\end{table}
	
	\section{Conclusions and Future Directives}
	Integrating IoT devices into everyday life requires efficient, timely data analysis
	through ML pipelines across cloud and edge computing. This research introduced
	the InTec framework, distributing the ML pipeline across Things, Edge, and Cloud
	layers to optimize computational efficiency. Through five experiments—focused on
	network traffic, latency, throughput, and power consumption—InTec demonstrated
	significant improvements: latency reduced by 81.56\%, network traffic decreased by
	10.92\%, and throughput improved by 9.82\%. Energy efficiency also increased, with
	21.86\% at the edge and 25.83\% at the cloud layers. These results highlight InTec’s
	transformative potential for scalable and resilient IoT systems.
	
	However, limitations remain, such as automating real-time algorithm selection,
	enhancing resilience through redundancy, and developing advanced monitoring tools are essential next steps. Cross-layer optimization, distributing computational
	tasks based on network conditions and device capabilities, will further boost efficiency and responsiveness in future iterations of InTec.
	
	The advancements and findings from the InTec framework open new horizons for
	research and development in IoT, edge computing, and ML. Future directions worth
	exploring include:
	
	\textit{Security and privacy concerns:} As InTec processes data at multiple layers
	(Things, Edge, and Cloud), it is vulnerable to various security and privacy risks.
	Future studies should investigate advanced encryption techniques, secure data transmission protocols, and decentralized trust mechanisms to safeguard sensitive data
	and maintain privacy throughout the ML pipeline.
	
	\textit{Interoperability with emerging technologies:} With the rapid advancement
	of IoT and AI technologies, it is crucial to ensure that InTec remains compatible
	with emerging standards and technologies (e.g., 5G/6G, AI accelerators, federated learning). Future research should explore how InTec can integrate and benefit from
	these advancements to enhance its efficiency and applicability further.
	
	\textit{Real-time adaptation to dynamic workloads:} Although InTec has improved efficiency, its ability to adapt to highly dynamic workloads and fluctuating network
	conditions remains limited. Enhancing InTec with real-time adaptation capabilities,
	where the framework can automatically adjust task distribution based on current
	conditions, is a promising direction for future work.
	
	\textit{Energy efficiency under resource constraints:} While InTec improves energy
	efficiency, its effectiveness in resource-constrained environments, such as battery-powered devices or Low-Power Wide-Area Networks (LPWANs), has not been fully
	explored. Investigating energy optimization strategies tailored for such scenarios
	could extend InTec’s applicability to a broader range of IoT devices and applications.
	
	\textit{Versatility across different domains and ML models:} Expanding InTec’s application beyond the MHEALTH dataset and incorporating other datasets will demonstrate its adaptability and performance across various ML problems. Future work
	will also explore using different ML models within the InTec framework to assess its
	versatility and ensure robustness and relevance in diverse real-world scenarios.
	
	\bibliographystyle{acm}
	\bibliography{InTec}

\begin{thebibliography}{10}

\bibitem{c16}
{\sc Achar, S., Faruqui, N., Whaiduzzaman, M., Awajan, A., and Alazab, M.}
\newblock Cyber-physical system security based on human activity recognition
  through {IoT} cloud computing.
\newblock {\em Electronics (Basel) 12}, 8 (Apr. 2023), 1892.

\bibitem{c33}
{\sc Arunachalam, M., Sanghavi, V., Kaira, S., and Ahuja, N.~A.}
\newblock End-to-end industrial {IoT}: Software optimization and acceleration.
\newblock {\em IEEE Internet Things M. 5}, 1 (Mar. 2022), 48--53.

\bibitem{c28}
{\sc Azar, J., Makhoul, A., Barhamgi, M., and Couturier, R.}
\newblock An energy efficient {IoT} data compression approach for edge machine
  learning.
\newblock {\em Future Gener. Comput. Syst. 96\/} (July 2019), 168--175.

\bibitem{c18}
{\sc Banos, O., Villalonga, C., Garcia, R., Saez, A., Damas, M.,
  Holgado-Terriza, J.~A., Lee, S., Pomares, H., and Rojas, I.}
\newblock Design, implementation and validation of a novel open framework for
  agile development of mobile health applications.
\newblock {\em Biomed. Eng. Online 14 Suppl 2}, Suppl 2 (Aug. 2015), S6.

\bibitem{c36}
{\sc Bianchi, V., Bassoli, M., Lombardo, G., Fornacciari, P., Mordonini, M.,
  and De~Munari, I.}
\newblock {IoT} wearable sensor and deep learning: An integrated approach for
  personalized human activity recognition in a smart home environment.
\newblock {\em IEEE Internet Things J. 6}, 5 (Oct. 2019), 8553--8562.

\bibitem{c35}
{\sc Bogacka, K., Sowi{\'n}ski, P., Danilenka, A., Biot, F.~M.,
  Wasielewska-Michniewska, K., Ganzha, M., Paprzycki, M., and Palau, C.~E.}
\newblock Flexible deployment of machine learning inference pipelines in the
  {Cloud--edge--IoT} continuum.
\newblock {\em Electronics (Basel) 13}, 10 (May 2024), 1888.

\bibitem{c11}
{\sc Chang, Z., Liu, S., Xiong, X., Cai, Z., and Tu, G.}
\newblock A survey of recent advances in edge-computing-powered artificial
  intelligence of things.
\newblock {\em IEEE Internet Things J. 8}, 18 (Sept. 2021), 13849--13875.

\bibitem{c10}
{\sc Chen, J., and Ran, X.}
\newblock Deep learning with edge computing: A review.
\newblock {\em Proc. IEEE Inst. Electr. Electron. Eng. 107}, 8 (Aug. 2019),
  1655--1674.

\bibitem{c34}
{\sc Fanariotis, A., Orphanoudakis, T., Kotrotsios, K., Fotopoulos, V.,
  Keramidas, G., and Karkazis, P.}
\newblock Power efficient machine learning models deployment on edge {IoT}
  devices.
\newblock {\em Sensors (Basel) 23}, 3 (Feb. 2023).

\bibitem{c24}
{\sc Ferlitsch, A.}
\newblock {\em Deep learning patterns and practices}.
\newblock Manning Publications, 2021.

\bibitem{c15}
{\sc Ghosh, A., and Grolinger, K.}
\newblock Edge-cloud computing for {IoT} data analytics: Embedding intelligence
  in the edge with deep learning.
\newblock {\em IEEE Trans. Industr. Inform.\/} (2020), 1--1.

\bibitem{c9}
{\sc Ghosh, A.~M., and Grolinger, K.}
\newblock Deep learning: Edge-cloud data analytics for {IoT}.
\newblock In {\em 2019 {IEEE} Canadian Conference of Electrical and Computer
  Engineering ({CCECE})\/} (2019), IEEE.

\bibitem{c3}
{\sc Gong, C., Lin, F., Gong, X., and Lu, Y.}
\newblock Intelligent cooperative edge computing in internet of things.
\newblock {\em IEEE Internet Things J. 7}, 10 (Oct. 2020), 9372--9382.

\bibitem{c23}
{\sc Hapke, H.}
\newblock {\em Building machine learning pipelines}.
\newblock O'Reilly Media, Sebastopol, CA, Aug. 2020.

\bibitem{c29}
{\sc Hu, L., Sun, G., and Ren, Y.}
\newblock {CoEdge}: Exploiting the edge-cloud collaboration for faster deep
  learning.
\newblock {\em IEEE Access 8\/} (2020), 100533--100541.

\bibitem{c30}
{\sc Janbi, N., Katib, I., Albeshri, A., and Mehmood, R.}
\newblock Distributed artificial intelligence-as-a-service ({DAIaaS}) for
  smarter {IoE} and {6G} environments.
\newblock {\em Sensors (Basel) 20}, 20 (Oct. 2020), 5796.

\bibitem{c31}
{\sc Kristiani, E., Yang, C.-T., Huang, C.-Y., Ko, P.-C., and Fathoni, H.}
\newblock On construction of sensors, edge, and cloud ({iSEC}) framework for
  smart system integration and applications.
\newblock {\em IEEE Internet Things J. 8}, 1 (Jan. 2021), 309--319.

\bibitem{c39}
{\sc Larian, H.}
\newblock \url{https://github.com/IDAS-Labratory/InTec_Framework}.
\newblock Accessed: 2025-2-8.

\bibitem{c19}
{\sc LeCun, Y., Bengio, Y., and Hinton, G.}
\newblock Deep learning.
\newblock {\em Nature 521}, 7553 (May 2015), 436--444.

\bibitem{c14}
{\sc Li, E., Zeng, L., Zhou, Z., and Chen, X.}
\newblock Edge {AI}: On-demand accelerating deep neural network inference via
  edge computing.
\newblock {\em IEEE Trans. Wirel. Commun. 19}, 1 (Jan. 2020), 447--457.

\bibitem{c25}
{\sc Li, L., Ota, K., and Dong, M.}
\newblock Deep learning for smart industry: Efficient manufacture inspection
  system with fog computing.
\newblock {\em IEEE Trans. Industr. Inform. 14}, 10 (Oct. 2018), 4665--4673.

\bibitem{c37}
{\sc Liang, F., Yu, W., Liu, X., Griffith, D., and Golmie, N.}
\newblock Towards edge-based deep learning in industrial internet of things.
\newblock {\em IEEE Internet Things J. 7}, 5 (May 2020), 4329--4341.

\bibitem{c38}
{\sc Lyu, L., Bezdek, J.~C., He, X., and Jin, J.}
\newblock Fog-embedded deep learning for the internet of things.
\newblock {\em IEEE Trans. Industr. Inform. 15}, 7 (July 2019), 4206--4215.

\bibitem{c27}
{\sc Manogaran, G., Shakeel, P.~M., Fouad, H., Nam, Y., Baskar, S.,
  Chilamkurti, N., and Sundarasekar, R.}
\newblock Wearable {IoT} smart-log patch: An edge computing-based bayesian deep
  learning network system for multi access physical monitoring system.
\newblock {\em Sensors (Basel) 19}, 13 (July 2019), 3030.

\bibitem{c8}
{\sc Merenda, M., Porcaro, C., and Iero, D.}
\newblock Edge machine learning for {AI-enabled} {IoT} devices: A review.
\newblock {\em Sensors (Basel) 20}, 9 (2020), 2533.

\bibitem{c32}
{\sc Raj, E., Buffoni, D., Westerlund, M., and Ahola, K.}
\newblock Edge {MLOps}: An automation framework for {AIoT} applications.
\newblock In {\em 2021 {IEEE} International Conference on Cloud Engineering
  ({IC2E})\/} (Oct. 2021), IEEE.

\bibitem{c20}
{\sc Razzaque, M.~A., Milojevic-Jevric, M., Palade, A., and Clarke, S.}
\newblock Middleware for internet of things: A survey.
\newblock {\em IEEE Internet Things J. 3}, 1 (Feb. 2016), 70--95.

\bibitem{c40}
{\sc Saez, Oresti~Banos, R.~G.}
\newblock {MHEALTH}, 2014.

\bibitem{c7}
{\sc Samie, F., Bauer, L., and Henkel, J.}
\newblock From cloud down to things: An overview of machine learning in
  internet of things.
\newblock {\em IEEE Internet Things J. 6}, 3 (2019), 4921--4934.

\bibitem{c22}
{\sc Satyanarayanan, M.}
\newblock The emergence of edge computing.
\newblock {\em Computer (Long Beach Calif.) 50}, 1 (Jan. 2017), 30--39.

\bibitem{c1}
{\sc Sharma, M., Tomar, A., and Hazra, A.}
\newblock Edge computing for industry 5.0: Fundamental, applications, and
  research challenges.
\newblock {\em IEEE Internet Things J. 11}, 11 (June 2024), 19070--19093.

\bibitem{c6}
{\sc Shi, Y., Yang, K., Jiang, T., Zhang, J., and Letaief, K.~B.}
\newblock Communication-efficient edge {AI}: Algorithms and systems.
\newblock {\em IEEE Commun. Surv. Tutor. 22}, 4 (2020), 2167--2191.

\bibitem{c21}
{\sc Sufyan, F., and Banerjee, A.}
\newblock Computation offloading for smart devices in fog-cloud queuing system.
\newblock {\em IETE J. Res.\/} (Jan. 2021), 1--13.

\bibitem{c17}
{\sc Wazwaz, A.~A., Amin, K.~M., Semari, N.~A., and Ghanem, T.~F.}
\newblock Enhancing human activity recognition using features reduction in
  {IoT} edge and azure cloud.
\newblock {\em Decision Analytics Journal 8}, 100282 (Sept. 2023), 100282.

\bibitem{c12}
{\sc Zhang, J., and Tao, D.}
\newblock Empowering things with intelligence: A survey of the progress,
  challenges, and opportunities in artificial intelligence of things.
\newblock {\em IEEE Internet Things J. 8}, 10 (May 2021), 7789--7817.

\bibitem{c26}
{\sc Zhao, Z., Barijough, K.~M., and Gerstlauer, A.}
\newblock {DeepThings}: Distributed adaptive deep learning inference on
  resource-constrained {IoT} edge clusters.
\newblock {\em IEEE Trans. Comput.-aided Des. Integr. Circuits Syst. 37}, 11
  (Nov. 2018), 2348--2359.

\bibitem{c13}
{\sc Zhou, Z., Chen, X., Li, E., Zeng, L., Luo, K., and Zhang, J.}
\newblock Edge intelligence: Paving the last mile of artificial intelligence
  with edge computing.
\newblock {\em Proc. IEEE Inst. Electr. Electron. Eng. 107}, 8 (Aug. 2019),
  1738--1762.

\end{thebibliography}

\end{document}